\documentclass[pre, twocolumn, aps, 10pt, a4paper, notitlepage, showkeys]{revtex4-1}

\usepackage{amsmath,amsfonts}
\usepackage[mathscr]{eucal}
\usepackage{graphicx}
\usepackage{color}
\usepackage{yfonts}
\usepackage{bm}
\usepackage{amssymb}
\usepackage{xspace}

\newcommand{\kB}{k_\mathrm{B}}                       
\newcommand{\Tc}{T_\mathrm{c}}                       
\newcommand{\cforce}{\mathfrak{F}_\mathrm{c}}        
\newcommand{\cforcez}{\mathfrak{F}_{\mathrm{c},z}}   
\newcommand{\cforcev}{\bm{\mathfrak{F}}_\mathrm{c}}  
\newcommand{\ctorquev}{\bm{\mathfrak{T}}_\mathrm{c}} 
\newcommand{\ctorquex}{\mathfrak{T}_{\mathrm{c},x}}  
\newcommand{\ctorquey}{\mathfrak{T}_{\mathrm{c},y}}  
\newcommand{\dd}{\mathrm{d}}                         
\newcommand{\fsc}{\mathscr{F}}                       
\newcommand{\fscv}{\bm{\fsc}}                        
\newcommand{\tsc}{\mathscr{T}}                       
\newcommand{\tscv}{\bm{\tsc}}                        
\newcommand{\fscslab}{\vartheta}                     
\newcommand{\pscslab}{\varphi}                       
\newcommand{\cpot}{\mathfrak{U}_\mathrm{c}}          
\newcommand{\psc}{\mathscr{U}}                       
\newcommand{\vctr}[1]{\bm{#1}}                       
\newcommand{\thp}{\theta_\mathrm{p}}                 
\newcommand{\rf}{\mathscr{O}}                        
\newcommand{\xib}{\xi_\mathrm{b}}                    
\newcommand{\sign}{\operatorname{sign}}              
\newcommand{\sphconf}{\Omega}                        
\newcommand{\degree}{^\circ}                         

\newcommand{\bsame}{\text{sm}}
\newcommand{\bopposite}{\text{op}}
\newcommand{\ampratio}{A_\xi}			     
\newcommand{\unit}[1]{\mathrm{#1}}                   
\newcommand{\Lambdapp}{\Lambda_{++}}                 
\newcommand{\Lambdapm}{\Lambda_{+-}}                 
\newcommand{\Lambdamp}{\Lambda_{-+}}                 
\newcommand{\Lambdamm}{\Lambda_{--}}                 
\newcommand{\Lambdasame}{\Lambda_{\text{sm}}}        
\newcommand{\Lambdaopposite}{\Lambda_{\text{op}}}    
\newcommand{\prdist}{\ell}     			     
\newcommand{\Heaviside}{\Theta_\mathrm{H}}           
\newcommand{\singi}{\textsf{I}\xspace}               
\newcommand{\singii}{\textsf{II}\xspace}             
\newcommand{\singiii}{\textsf{III}\xspace}           
\newcommand{\singiv}{\textsf{IV}\xspace}             
\newcommand{\tmpvari}{\hat{\varepsilon}}             
\newcommand{\tmpvarii}{\sigma}                       
\newcommand{\spdim}{d}				     
\newcommand{\spomega}{\hat{\omega}}                  

\begin{document}
\title{Effective Pair Interaction of Patchy Particles in Critical Fluids}

\author{N. Farahmand Bafi}
\email{nimabafi@is.mpg.de}
\affiliation{Max--Planck--Institut f\"ur Intelligente Systeme, Heisenbergstr.~3, 70569 Stuttgart, Germany}
\affiliation{Institut f\"ur Theoretische Physik IV, Universit\"at Stuttgart, Pfaffenwaldring 57, 70569 Stuttgart, Germany}
\author{P. Nowakowski}
\email{pionow@is.mpg.de}
\affiliation{Max--Planck--Institut f\"ur Intelligente Systeme, Heisenbergstr.~3, 70569 Stuttgart, Germany}
\affiliation{Institut f\"ur Theoretische Physik IV, Universit\"at Stuttgart, Pfaffenwaldring 57, 70569 Stuttgart, Germany}
\author{S. Dietrich}
\affiliation{Max--Planck--Institut f\"ur Intelligente Systeme, Heisenbergstr.~3, 70569 Stuttgart, Germany}
\affiliation{Institut f\"ur Theoretische Physik IV, Universit\"at Stuttgart, Pfaffenwaldring 57, 70569 Stuttgart, Germany}

\begin{abstract}
\begin{center}
\textbf{Abstract}
\end{center}

We study the critical Casimir interaction between two spherical colloids immersed in a binary liquid mixture close to its critical demixing point. The surface of each colloid prefers one species of the mixture with the exception of a circular patch of arbitrary size, where the other species is preferred. For such objects we calculate, within the Derjaguin approximation, the~scaling function describing the critical Casimir potential, and we use it to derive the scaling functions for all components of the forces and torques acting on both colloids. The results are compared with available experimental data. Moreover, the general relation between the scaling function for the~potential and the scaling functions for the force and the torque is derived.
\end{abstract}

\keywords{critical Casimir interaction, colloidal particles, Derjaguin approximation}

\maketitle

\section{Introduction}\label{secA}


Colloids have been the subject of research for centuries \cite{Brown1828, Einstein1905, Perrin1909, Lekkerkerker2011}. The initial studies were mostly concerned with the observation and explanation of the behavior of naturally occurring colloids in suspensions, as they are small enough to exhibit certain properties typical for molecular systems and, at the same time, they are big enough to be directly observable by using a microscope. With the development of the corresponding theoretical description \cite{Pawar2010, Liang2007, Lekkerkerker2011} and methods of synthesis \cite{Pawar2010,Yi2013} it has become possible to design colloidal particles exhibiting desired properties; this has found applications in various areas like pattern formations \cite{Pawar2010,Yi2013}, drug delivery~\cite{Bunjes2010, Fanun2016}, phoretic motors~\cite{Popescu2018}, in the oil industry~\cite{Toulhoat1992}, and numerous others~\cite{Caruso2006}. In many cases, the description of the system can be simplified by introducing an effective interaction between the colloids, which is mediated by the solvent and is present in addition to their direct interaction. Different types of such effective interactions have been proposed~\cite{Liang2007, Lekkerkerker2011} like London--van der Waals forces, screened electrostatic repulsion, steric, and depletion forces. One of the ways of inducing and controlling the interaction between colloids is the use of the critical Casimir effect~\cite{Fisher1978, Maciolek2018}.

The critical Casimir force is one of the manifestations of effective interactions induced by fluctuations. Historically, the first example of recognizing such a force was the electromagnetic Casimir effect \cite{Casimir1948}, according to which the force acting between two conductors originates from the confinement induced restrictions of the quantum fluctuations of the electromagnetic field. In the case of the  critical Casimir effect, two or more objects are immersed in a medium, in which the thermodynamic state is tuned to be close to its critical point. For these latter systems, the order parameter fluctuations are large, and the resulting effective interaction between colloids is long--ranged~\cite{Fisher1978}.

One of the most interesting features of critical phenomena is the concept of universality~\cite{Kadanoff1976}, which stipulates that in the vicinity of the critical point certain properties of the system (such as critical indices, ratios of amplitudes, and scaling functions) depend only on certain general features, like the spatial dimension and the number of components of the order parameter, but not on the microscopic details of the system. This allows one to group specific critical systems into various so--called bulk universality classes, which provides a convenient means of theoretical analysis: instead of modeling a complicated system one can study a much simpler model which can act as a representative of the corresponding universality class. One of the prominent examples is the 3D Ising universality class which contains a simple fluid close to its critical point, a uniaxial ferromagnet in the vicinity of the Curie point, and a binary liquid mixture close to its critical demixing point~\cite{Brankov2000,Krech1994}.

In the case of semi--infinite systems, each bulk universality class splits into several surface universality classes, depending on some general properties of the interaction between the critical system and the confining wall. Similarly, for systems with a slab geometry there emerge various film universality classes which can typically be characterized by pairs of surface universality classes of the two confining walls. The universality of the critical Casimir force manifests itself via the occurrence of scaling laws: close to the critical point the force can be expressed in terms of a power law times a universal scaling function. The latter depends only on dimensionless ratios of geometrical parameters describing the system, the bulk correlation length, and suitable scaled bulk fields. Universality of the scaling functions implies that they are identical for systems from the same bulk and film universality classes~\cite{Brankov2000, Krech1994}.

Early studies of the critical Casimir force were solely theoretical and focused on the slab geometry \cite{Fisher1978,Krech1992a,Krech1992b}, but they have soon been extended to spheres~\cite{Burkhardt1995, *Burkhardt1997, Eisenriegler1995, Hanke1998, Schlesener2003, Gambassi2009a, Troendle2010, Dang2013, Mattos2013, Hasenbusch2013, Mohry2014, Vasilyev2014, Edison2015, Hobrecht2015, Labbe2016, Tasios2016}, ellipsoids \cite{Kondrat2007, Kondrat2009}, and more complicated systems~\cite{Emig2003, Vasilyev2013, Labbe2014, Bimonte2015, Maciolek2015,Troendle2015}.
These predictions were verified experimentally in the context of wetting films~\cite{Garcia1999, Mukhopadhyay1999, Mukhopadhyay2001, Garcia2002, Ganshin2002, Fukuto2005, Rafai2007} and colloidal suspensions~\cite{Hertlein2008, Soyka2008, Bonn2009, Troendle2011, Veen2012, Shelke2013, Nguyen2013, Paladugu2016}.
In the latter case, it was demonstrated that critical Casimir forces --- due to their temperature sensitivity --- can provide a means of inducing and controlling self--assembly and structure formation~\cite{Nguyen2016, Nguyen2017}.

Recent advances in synthesis have facilitated the fabrication of colloidal particles in a controlled way with spatially varying surface properties. Since the effective interactions between such particles are anisotropic, the resulting self--assembly patterns can be much more complex than in the case of chemically homogeneous particles~\cite{deGennes1992,Pawar2010,Yi2013,Maciolek2018,Oh2019}. It was observed that the properties of critical Casimir interactions allow one to change the self--assembly structure of certain colloids by varying the thermodynamic parameters of the solvent such as temperature and concentration~\cite{Nguyen2016, Nguyen2017}. These observations have also been confirmed by various numerical simulations of chemically inhomogeneous particles~\cite{Tasios2016, Maciolek2018}.

A full understanding of the relation between the properties of a single colloidal particle and the pattern formed by a large number of such particles can potentially provide a useful tool to create any kind of three--dimensional microstructures. In this respect, one of the necessary steps is to investigate the critical Casimir pair interaction for inhomogeneous particles. Some theoretical studies \cite{Sprenger2005, Sprenger2006, Troendle2009, Labbe2014, Labbe2016, Nowakowski2016} have already addressed this issue by using mean field theory \cite{Weiss1907}, the Derjaguin approximation \cite{Derjaguin1934}, and the exact two--dimensional solution \cite{Onsager1944}. So far, these studies were devoted to either patterned surfaces or particles with distinct chemical properties on half of their surface (so--called Janus particles \cite{deGennes1992}).

In the present contribution, we extend the studies of equilibrium critical Casimir interactions to the case of two identical spherical colloids with a circular cap forming a chemical patch of arbitrary size on their surfaces. We use the Derjaguin approximation in order to determine all components of the force and the torque acting on them. These calculations can straightforwardly be generalized to more complicated chemical patterns. We compare our results with available experimental data.

The paper is organized as follows: In Sec.~\ref{secB} we introduce spherical colloids with chemically inhomogeneous surfaces. Section~\ref{secC} is devoted to critical Casimir effect; we recall the pertinent results for the slab geometry and for two spheres. Moreover we discuss the Derjaguin approximation used in our calculations. In Sec.~\ref{secD} the procedure of calculating forces and torques acting on the colloidal particles is described. Afterwords, in Sec.~\ref{secE} we comment on the validity and accuracy of the approximation used in our calculations. Next, in Sec.~\ref{secK} we present our numerical results and compare them with available experimental data. Finally, the summary of our research is presented in Sec.~\ref{secU}. Our presentation is supplemented by three appendices. In Appendix~\ref{secV} we derive the formulae relating the scaling function for the critical Casimir potential with the scaling functions for the components of the critical Casimir force and torque; in Appendix~\ref{secX} we recall the scaling functions for the critical Casimir force in the slab geometry; and in Appendix~\ref{secW} we discuss nonanalyticities of the scaling functions.


\section{Patchy particles}\label{secB}


We consider a system of two spherical colloidal particles immersed in a binary liquid mixture close to its critical demixing point. We assume that the mixture consists of two species A and B, the composition of which is equal to the one of the critical point, and that the temperature $T$ is close to the critical one $\Tc$. The two colloidal particles are spheres of the same radius $R$ and the surface--to--surface distance between them is denoted as $D$. In this study we assume that the forces and torques acting between the colloids are balanced by external forces such that the particles are kept in fixed positions and the system is in thermodynamic equilibrium. We do not consider any dynamic effects. 

The surface of the particles is inhomogeneous, i.e., the interaction with the two components A and B of the mixture depends on the position on the surface. We study the critical Casimir interaction in the scaling limit (cf.~Sec. \ref{secC}), in which only general properties of the wall--fluid interaction are relevant. In order to describe the surface it is sufficient to specify at each point which component of the mixture is preferred. In all figures, the regions where the component A of the mixture is preferred are denoted by `$+$' and plotted in red while the preference for component B is denoted by `$-$' and plotted in blue. The detailed interaction between the mixture and the surface of the spheres gives rise to subdominant terms in the scaling limit, i.e., corrections to scaling; studying them is beyond the scope of the present analysis.

We note that the approach used here renders all crossovers between regions of different affinity of the surface to be sharp. A more realistic, gradual description of such interfaces calls for a separate study.

\begin{figure}
 \includegraphics[width=0.45\textwidth]{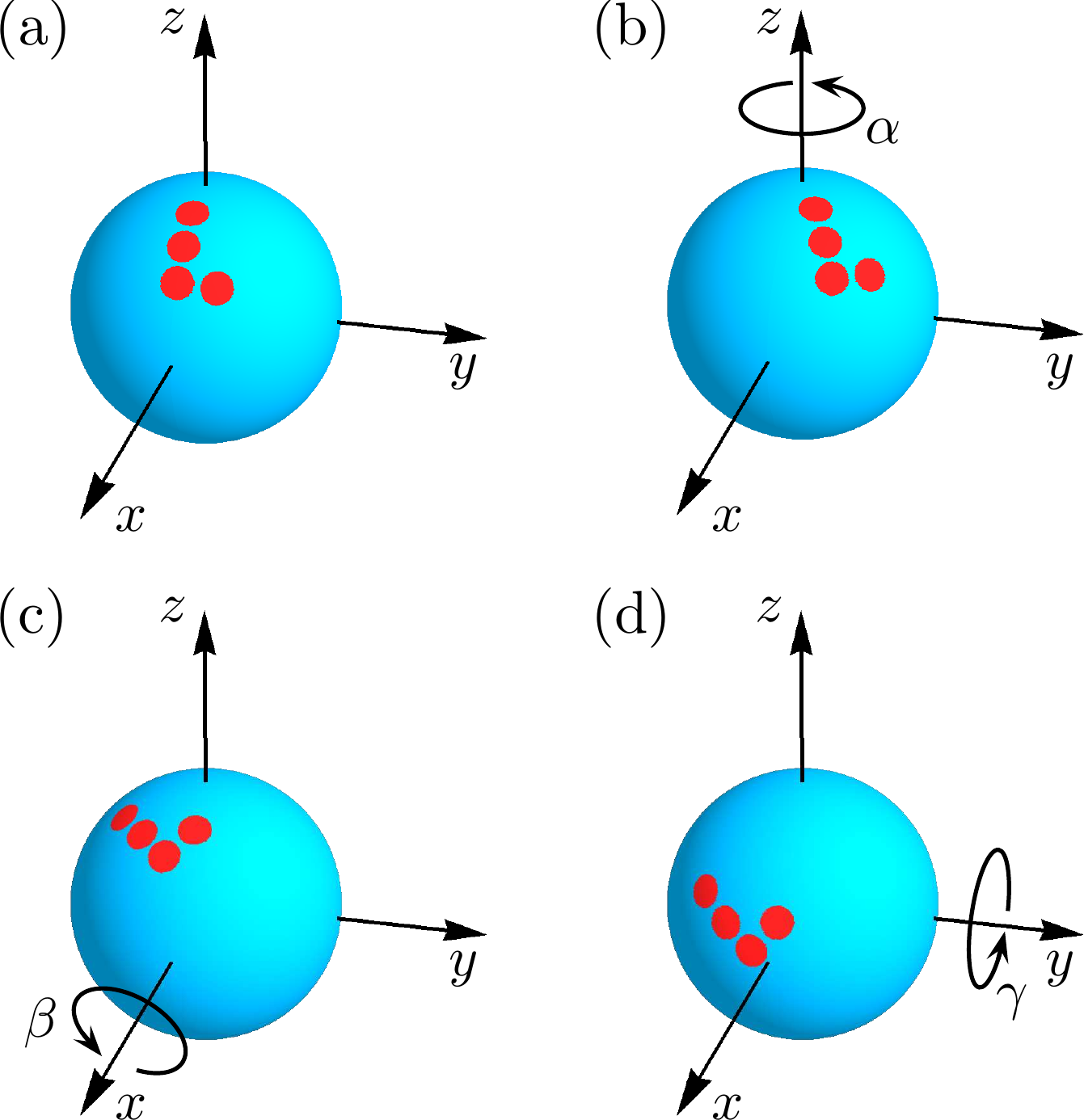}
 \caption{\label{secB:fig1} Schematic plot of the three rotations which define the rotational configuration of the colloid. The particle with an arbitrary pattern (a) is, firstly, rotated by the angle $\alpha$ around the $z$ axis (b); secondly, it is rotated by the angle $\beta$ around the $x$ axis (c); and, thirdly, it is rotated by the angle $\gamma$ around the $y$ axis (d). The pattern shown in this picture is chosen in order to illustrate all the transformations, but it is not studied in the present analysis.}
\end{figure}

In order to fully describe the configuration of the system, we assume that the center of the first colloid is located at the origin of the laboratory reference frame $\rf$, and the center of the second one is at the point defined by a vector $\vctr{r}$ of length $D+2R$. The rotational configuration of each colloid is determined by three angles $\alpha, \beta,$ and $\gamma$ in accordance with the following procedure: The colloid is initially put with its center at the origin of $\rf$ in a predefined initial configuration. First, it is rotated around the $z$ axis by the angle $\alpha$. The second rotation is by the angle $\beta$ around the $x$ axis, and the third one is by the angle $\gamma$ around the $y$ axis. Finally, in order to obtain the desired configuration, the second colloid is shifted by the vector~$\vctr{r}$. All three rotations are active and in the direction determined by the right--hand rule. The procedure is shown schematically in Fig.~\ref{secB:fig1}. All applied rotations are represented by the matrix
\begin{widetext}
\begin{equation}\label{secB:stdrotation}
 \mathbb{R}\left(\alpha,\beta,\gamma\right)=\begin{pmatrix}
             & & & & \vspace{-0.65cm}\\
             \cos\alpha\,\cos\gamma+\sin\alpha\,\sin\beta\,\sin\gamma & \vdots &-\sin\alpha\,\cos\gamma+\cos\alpha\,\sin\beta\,\sin\gamma & \vdots & \cos\beta\,\sin\gamma \vspace{-0.25cm} \\ \vspace{-0.25cm}
             \dotfill & \!\!\!\ldots\!\!\! &\dotfill & \!\!\!\ldots\!\!\! & \dotfill \\
             \sin\alpha\,\cos\beta & \vdots & \cos\alpha\,\cos\beta & \vdots & -\sin\beta \vspace{-0.25cm} \\ \vspace{-0.25cm}
             \dotfill & \!\!\!\ldots\!\!\! &\dotfill& \!\!\!\ldots\!\!\!  & \dotfill \\
             \sin\alpha\,\sin\beta\,\cos\gamma-\cos\alpha\,\sin\gamma & \vdots &\cos\alpha\,\sin\beta\,\cos\gamma+\sin\alpha\,\sin\gamma &\vdots & \cos\beta\,\cos\gamma
            \end{pmatrix}.\vspace{-0.3cm}
\end{equation}
\end{widetext}

In order to obtain any possible rotational configuration, it is sufficient to consider $\alpha\in\left[0\degree,360\degree\right)$, $\beta\in\left(-90\degree, 90\degree\right]$, and $\gamma\in\left[0\degree,360\degree\right)$. The case of $\beta= 90\degree$ requires special care, because in this case the rotations around the $z$ axis and the $y$ axis are, in fact, the same rotation and thus one can assume that $\gamma=0$. We note that it is often helpful to consider $\alpha$, $\beta$, or $\gamma$ beyond the domains given above; in such cases the resulting rotations can always be replaced by the ones which fulfill the constraints.

It is convenient to introduce
\begin{equation}
 \sphconf=\left(\alpha_1, \beta_1, \gamma_1, \alpha_2, \beta_2, \gamma_2\right),
\end{equation}
in order to denote the rotational configuration of the colloids. Here, $\alpha_i$, $\beta_i$, and $\gamma_i$ describe the configuration of the first ($i=1$) and of the second ($i=2$) particle. Accordingly, the system is described by four quantities: temperature $T$, radius $R$ of the particles, relative position $\vctr{r}$ of the colloids, and the rotational configuration $\sphconf$. The surface--to--surface distance follows from the relation $D=\left|\vctr{r}\right|-2R$.

In order to study the system of two colloids it is not necessary to consider all possible configurations, because the critical Casimir interaction is invariant under rotations. (The translational symmetry has already been utilized by keeping the first particle at the origin.) We introduce the rotation $\mathbb{T}$ which moves the center of the second particle to the positive $y$ semi--axis, i.e.,
\begin{equation}\label{secB:T}
 \mathbb{T}\, \vctr{r}=r\,\vctr{e}_y,
\end{equation}
where $\vctr{e}_y$ is the unit vector in $y$ direction. With such a rotation, not only the vector $\vctr{r}$ is transformed, but also the rotational configuration $\sphconf$; we denote the new configuration by
\begin{equation}
 \sphconf^\mathbb{T}=\left(\alpha_1^\mathbb{T}, \beta_1^\mathbb{T}, \gamma_1^\mathbb{T}, \alpha_2^\mathbb{T}, \beta_2^\mathbb{T}, \gamma_2^\mathbb{T}\right).
\end{equation}
We note that $\mathbb{T}$ is not defined uniquely by the relation in Eq.~\eqref{secB:T}; composing $\mathbb{T}$ with any rotation around the $y$ axis preserves Eq.~\eqref{secB:T}. Since the additional rotation of the colloids around the $y$ axis is only increasing $\gamma_1^\mathbb{T}$ and $\gamma_2^\mathbb{T}$ by the angle of the rotation, it is possible to choose $\mathbb{T}$ in such a way, that $\gamma_2^\mathbb{T}=0$. This additional condition renders $\mathbb{T}$ unique. Concerning an example of constructing the matrix $\mathbb{T}$ see the calculation presented in Appendix~\ref{secV}.

\begin{figure}
 \includegraphics[width=0.45\textwidth]{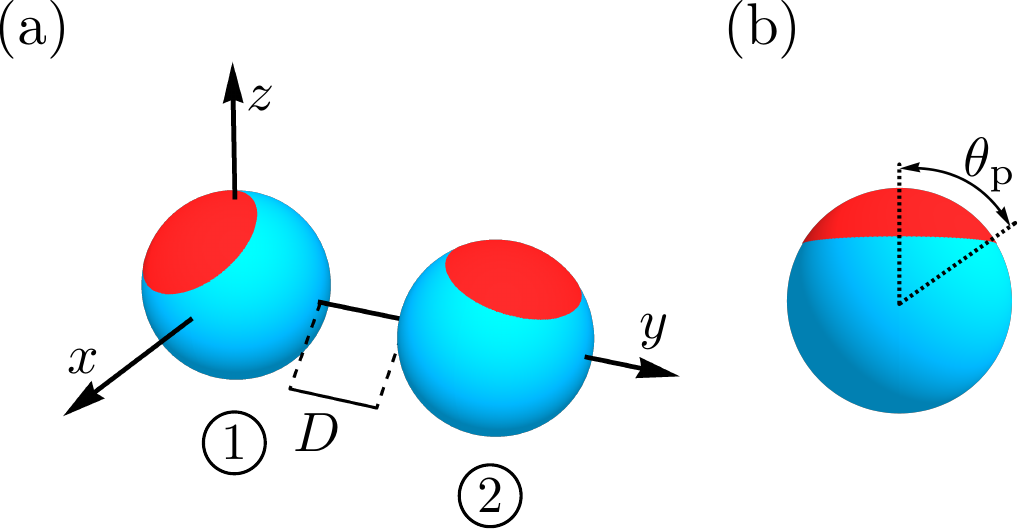}
  \caption{\label{secB:fig2}(a)~Schematic plot of two spherical colloids in a special configuration (see the main text). In this configuration the center of the first particle is located at the origin and the center of the second particle is on the positive part of the $y$ axis. The surface--to--surface distance between the colloids is denoted as $D$. (b)~Schematic plot of the spherical particle with a single circular patch. This
  type of particles is the one considered in all subsequent calculations. The angle $\thp$ defines the size of the patch.}
\end{figure}

We call the configurations of colloids, for which $\vctr{r}=r\,\vctr{e}_y$ and $\gamma_2=0$, \textit{special configurations}. As we have shown above, any general configuration of two patchy particles can be transformed to the special one by the rotation $\mathbb{T}$ constructed above. 
Therefore, it is sufficient to calculate the critical Casimir forces and torques only for the special configurations. An example of such a configuration is presented in Fig.~\ref{secB:fig2}(a). We note that, because forces and torques are vectors, in order to calculate them for an arbitrary configuration one must transform the results obtained for the special configuration by the rotation $\mathbb{T}^{-1}$.

For any configuration of the colloids we additionally introduce the \textit{relative rotational configuration}
\begin{equation}\label{secB:Omegaast}
\sphconf^\ast=\left(\alpha_1^\ast, \beta_1^\ast, \gamma_1^\ast, \alpha_2^\ast, \beta_2^\ast\right)
\end{equation}
as the configuration the particles would have if they were transformed to the special configuration. All angles in $\sphconf^\ast$ are (rather complicated) functions of $\vctr{r}$ and $\sphconf$.

Wherever possible, we introduce the laboratory reference frame $\rf$ such that the particles already assume the special configuration. This way, we do not have to determine the rotation matrix $\mathbb{T}$ which simplifies the calculation.

The above discussion is valid for an arbitrary pattern on the surfaces of the colloids (and can easily be generalized to nonspherical particles). Here, we consider only a simple pattern with one circular patch on the surface of each sphere: in the initial position of the particle (before applying any of the rotations), the component A (B) of the mixture is preferred for $\theta\leqslant\thp$ ($\theta>\thp$) where $\theta$ is the polar angle of the spherical coordinates in $\rf$, and the opening angle $\thp$ is the parameter describing the angular size of the circular patch. A schematic plot of the colloid is presented in Fig.~\ref{secB:fig2}(b). For such a pattern the first rotation (by an angle $\alpha$ around the $z$ axis) does not change the configuration and thus $\sphconf$ is defined completely by $\beta_1$, $\gamma_1$, $\beta_2$, and $\gamma_2$.

The critical Casimir interaction in the special case $\thp=90\degree$ has already been addressed in Ref.~\cite{Labbe2016}. This corresponds to a so--called \textit{Janus particle} \cite{deGennes1999, Hu2012}, in which the particle consists of two hemispheres preferring opposite components of the binary liquid mixture.

\section{Critical Casimir force}\label{secC}

In order to calculate the critical Casimir interaction between colloids we use the Derjaguin approximation which is based on the slab geometry. In this section, we recall all pertinent results for the slab and the spherical geometries and adapt them to the present case of chemically inhomogeneous surfaces.

\subsection{Slab geometry}\label{secC:sg}

We start from the description of the thermodynamic state of the binary liquid mixture. In general, such a liquid is fully described by three intensive parameters. In the current study we assume that the pressure $p$ is fixed and the concentrations (molar fractions $x_\text{A}$ and $x_\text{B}=1-x_\text{A}$) of the components of the mixture are tuned to be equal to the critical ones. The temperature $T$, as the third parameter, is free to change. We assume that it is close to the critical temperature $\Tc$. (The values of both the critical concentration and the critical temperature depend on the pressure $p$.) If such a liquid is confined by two macroscopically large, parallel walls, the resulting slab system is described by only two macroscopic control parameters: the temperature $T$ and the distance $L$ between the walls.


Close to the critical point, the critical fluctuations of the concentration of the fluid lead to the effective critical Casimir force \cite{Fisher1978} acting between the walls. In spatial dimension $\spdim=3$ this force can be described by the universal scaling formula
\begin{equation}\label{secC:cforceslab}
 \cforce^\mathrm{slab}\left(L,T\right)/A=\frac{\kB \Tc}{L^3}\fscslab_{\mathrm{s}}\left(\omega\right),
\end{equation}
where $\kB$ is the Boltzmann constant, $\cforce^\mathrm{slab}/A$ is the critical Casimir force per area (i.e., excess pressure); $\fscslab_\mathrm{s}\left(\omega\right)$ is a scaling function which is universal, i.e., it is independent of the microscopic details of the system. The scaling function depends only on the bulk universality class of the critical point of the fluid, and on the interaction between the fluid and the two walls, encoded into the corresponding film universality class~\cite{Krech1994}, denoted by the index `$\mathrm{s}$' (see below). The scaling variable is
\begin{equation}\label{secC:scalingvariable}
 \omega=\frac{L}{\xi_0^\pm \left|t\right|^{-\nu}} \sign\left(t\right),
\end{equation}
where $\xib(t\rightarrow 0^{\pm})=\xi_0^{\pm}|t|^{-\nu}$ is the bulk correlation length, $\xi_0^+$ ($\xi_0^-$) is its amplitude for $T>\Tc$ ($T<\Tc$) in the case of an upper critical point of a binary liquid mixture, $t=\left(T-\Tc\right)/\Tc$ is the reduced temperature, and $\nu$ is the critical exponent of the correlation length.

The expression in Eq.~\eqref{secC:cforceslab} is exact in the scaling limit, i.e., for $T\to\Tc$ and $L\to\infty$ with $\omega$ fixed. If $T$ is fixed and close to $\Tc$, and $L$ is large but finite, Eq.~\eqref{secC:cforceslab} provides an approximation of the actual critical Casimir pressure; in order to improve the result, one has to include corrections to scaling (such as higher order terms in $1/L$).

Here, we consider binary liquid mixtures exhibiting a critical demixing point, which belongs to the universality class of the 3D Ising model, so that \cite{Pelissetto2002}
\begin{equation}\label{secC:ampratio}
 \nu=0.6301(4),\quad \ampratio=\xi_0^+/\xi_0^-=1.896(10).
\end{equation}
In the present context we are interested only in the surface universality class of a symmetry breaking surface field in which the surface prefers either A (`$+$') or B (`$-$') chemical species of the binary liquid mixture. For our study only two film universality classes are relevant: If both walls prefer the same component of the mixture the scaling function in Eq.~\eqref{secC:cforceslab} is $\fscslab_\bsame\left(\omega\right)$ ($\mathrm{s}=\mathrm{sm}$, same boundary conditions: `$++$' or `$--$') and if the walls prefer different components of the binary mixture it is $\fscslab_\bopposite\left(\omega\right)$ ($\mathrm{s}=\mathrm{op}$, opposite boundary conditions: `$+-$' or `$-+$'). Since the analytical forms of these scaling functions are not known, we use their numerical estimates in spatial dimension $\spdim=3$ (see, c.f.,~Sec.~\ref{secD:sfslab} and Fig.~\ref{secD:Slab-Scaling}).

Finally, we recall the scaling formula for the potential of the critical Casimir force:
\begin{equation}\label{secC:pscslabdef}
 \cpot^\text{slab}\left(L,T\right)/A=\frac{\kB \Tc}{L^2}\, \pscslab_\mathrm{s}\left(\omega\right),
\end{equation}
where $\pscslab_\mathrm{s}\left(\omega\right)$ is a universal scaling function 
with $\mathrm{s}=\mathrm{sm}$ or $\mathrm{s}=\mathrm{op}$; the form of this function can be determined from $\fscslab_\mathrm{s}\left(\omega\right)$ by using the relation 
\begin{equation}
 \pscslab_\mathrm{s}\left(\omega\right)=
 \begin{cases}
   \omega^2\displaystyle\int_{\omega}^\infty \dd \zeta\ \fscslab_\mathrm{s}\left(\zeta\right)/\zeta^3,& \omega>0,\\
   \displaystyle\fscslab_\mathrm{s}\left(0\right)/2, & \omega=0,\\
   -\omega^2\displaystyle\int^{\omega}_{-\infty} \dd \zeta\ \fscslab_\mathrm{s}\left(\zeta\right)/\zeta^3,& \omega<0,
 \end{cases}
\end{equation}
which follows directly from the relation between the critical Casimir force and its potential.

\subsection{Spherical objects}

When two spherical colloids are immersed into a critical fluid, like in the slab geometry, fluctuations induce a critical Casimir interaction between them. For homogeneous spheres, this effect has been studied theoretically \cite{Burkhardt1995, *Burkhardt1997, Hanke1998}, numerically \cite{Vasilyev2009a, *Vasilyev2009b}, and experimentally \cite{Bonn2009}. There are three macroscopic parameters describing the system: the temperature $T$, the radius $R$ of the colloids, and the surface--to--surface distance $D$ between them. Following the literature, these variables are combined into the following two scaling variables:
\begin{equation}\label{secC:DeltaTheta}
 \Delta=\frac{D}{R}, \quad \Theta= \frac{D\operatorname{sign}\left(t\right)}{\xi_0^\pm \left|t\right|^{-\nu}}.
\end{equation}

If the spheres are chemically homogeneous, due to symmetry the critical Casimir force acts in radial direction only. Close to the critical point, it is given by the scaling law
\begin{equation}\label{secC:ccfhspheres}
 \cforce^{\mathrm{H}}\left(T,R,D\right)=\frac{\kB \Tc}{R} \frac{\fsc^{\mathrm{H}}_\mathrm{s}\left(\Delta,\Theta\right)}{\Delta^2},
\end{equation}
where the index `$\mathrm{H}$' indicates that the considered quantity is evaluated for homogeneous spheres, $\fsc^{\mathrm{H}}_\mathrm{s}$ is a universal scaling function where the index $\mathrm{s}=\text{sm}$ or $\mathrm{s}=\text{op}$ denotes the same or opposite affinities of the two spheres, respectively. The additional factor $\Delta^{-2}$ has been introduced in order to render the scaling function finite in the limit $\Delta\to 0$ \cite{Burkhardt1995, *Burkhardt1997}. The potential of the critical Casimir force in the homogeneous case is given by
\begin{equation}
\label{secC:ccpotSphere}
 \cpot^{\mathrm{H}}\left(T,R,D\right)=\kB \Tc \frac{\psc^{\mathrm{H}}_\mathrm{s}\left(\Delta,\Theta\right)}{\Delta},
\end{equation}
where $\psc^{\mathrm{H}}_\mathrm{s}$ is another universal scaling function. Like in the slab geometry, the scaling functions $\fsc^{\mathrm{H}}_\mathrm{s}$ and $\psc^{\mathrm{H}}_\mathrm{s}$ are related. Equations~\eqref{secC:ccfhspheres} and \eqref{secC:ccpotSphere} are valid in the scaling limit $T\to\Tc$, $D\to\infty$, $R\to\infty$ with $\Delta$ and $\Theta$ fixed.

We now turn to the case of inhomogeneous colloids studied here. If the preferences for the two components of the binary mixture vary along the colloid surface, the critical Casimir interaction is modified relative to the homogeneous case; in general, the force becomes non--radial and a torque appears.

In the special configuration (see Sec.~\ref{secB}), Eq.~\eqref{secC:ccfhspheres} can be generalized to the following scaling formulae:
\begin{subequations}\label{secC:FTscaling}
\begin{align}
 \cforcev^{\left(i\right)} \left(T,R,D,\sphconf^\ast\right)&=\frac{\kB \Tc}{R}\frac{\fscv^{\left(i\right)}\left(\Delta, \Theta, \sphconf^\ast\right)}{\Delta^2} \label{secC:forcesf}\\
 \ctorquev^{\left(i\right)}\left(T,R,D,\sphconf^\ast\right)&=\kB \Tc\frac{\tscv^{\left(i\right)}\left(\Delta, \Theta, \sphconf^\ast\right)}{\Delta^2}, \label{secC:torquesf}
\end{align}
\end{subequations}
where $\fscv^{\left(i\right)}$ and $\tscv^{\left(i\right)}$ denote the vector scaling functions for the force $\cforcev^{\left(i\right)}$ and the torque $\ctorquev^{\left(i\right)}$ acting on the inhomogeneous spheres; $i=1,2$ labels the particles; $\sphconf^\ast$ denotes the relative orientation of the colloids (see Eq.~\eqref{secB:Omegaast}); and the scaling variables $\Delta$ and $\Theta$ are given by Eq.~\eqref{secC:DeltaTheta}. Note that we consider here the torques to be acting on the center of each particle.

Since the system is in thermal equilibrium, the total force and torque must vanish. This allows one to relate the above scaling functions:
\begin{subequations}\label{secC:FTrelations}
\begin{align}
 \label{secC:Frelation}\fscv^{\left(1\right)}+\fscv^{\left(2\right)}&=\vctr{0},\\
 \label{secC:Trelation}\tscv^{\left(1\right)}+\tscv^{\left(2\right)}+\left(2+\Delta\right)\vctr{e}_y\times \fscv^{\left(2\right)}&=\vctr{0}.
\end{align}
\end{subequations}
The formulae show that it is sufficient to calculate the force and the torque acting on one particle; the other quantities follow from Eq.~\eqref{secC:FTrelations}.

The critical Casimir potential of interaction between the colloids can also be determined in terms of an appropriate scaling law:
\begin{equation}\label{secC:potscaling}
 \cpot\left(T, R, D, \sphconf^\ast\right)=\kB \Tc \frac{\psc\left(\Delta, \Theta, \sphconf^\ast\right)}{\Delta},
\end{equation}
where $\psc$ is the scaling function and the factor $\Delta^{-1}$ has been split off in order to keep the scaling function finite in the limit $\Delta\to 0$. Unlike force and torque, the potential is a scalar quantity and thus the scaling formula in Eq.~\eqref{secC:potscaling} holds even if the second colloid is not located on the $y$ axis (because $\sphconf^\ast$ is the relative configuration of the spheres, it is invariant under the rotations of the system). The relation between the scaling function for the potential and the scaling functions for the forces and torques is derived in Appendix~\ref{secV} and is given by Eq.~\eqref{secV:forcetorque}.

Finally, we note that the scaling laws in Eqs.~\eqref{secC:FTscaling} and \eqref{secC:potscaling}, together with the relation between the scaling functions studied in Appendix~\ref{secV}, are very general and can be used for particles of arbitrary shapes and surface patterns.

\subsection{Derjaguin approximation}\label{secC:Derjaguin}

\begin{figure}
\includegraphics[width=0.45\textwidth]{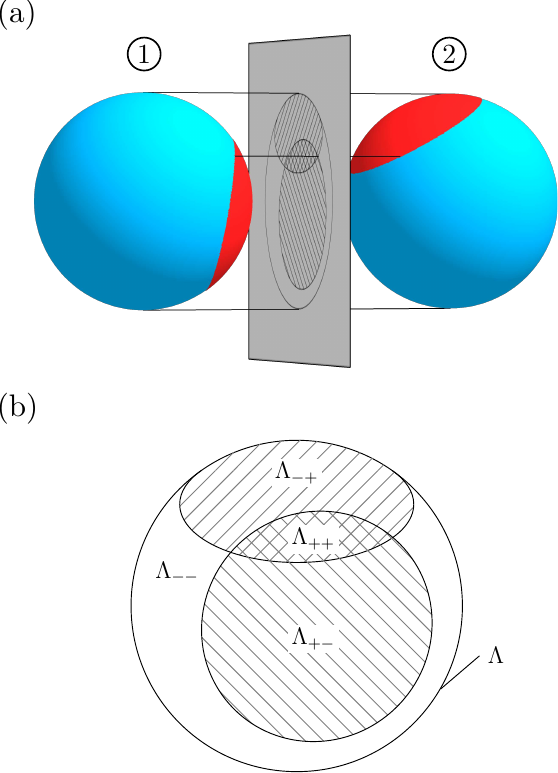} 
\caption{Schematic plot of the (grayish) projection plane used in the Derjaguin approximation for an exemplary configuration of particles. Within this approximation the interaction between spheres depends only on points from the right hemisphere of the first particle and the left hemisphere of the second particle. Panel (a) illustrates the process of the orthogonal projection of the interacting surfaces (facing each other) to a plane parallel to the $x$ and $z$ axes which is the projection plane. The horizontal lines illustrate the projection for three points on the projection plane serving as examples. Panel (b) presents the resulting pattern on the projection plane. The projection of both spheres gives the same circle $\Lambda$ of radius $R$, which we divide into four disjoint (nonoverlapping) regions $\Lambda_{++}$, $\Lambda_{+-}$, $\Lambda_{-+}$, and $\Lambda_{--}$. The first sign in the index of $\Lambda$ denotes the affinity of the surface of the first particle and the second sign of the second particle. The symbol `$+$' means that there is a patch at an appropriate point of the surface of the particle (denoted with red color) and `$-$' denotes that there is no patch (blue color).}
\label{secC:pplane}
\end{figure}

In order to obtain the scaling functions for the critical Casimir interaction one can use several distinct techniques like mean field theory \cite{Weiss1907}, Monte Carlo simulations \cite{Metropolis1953}, or the so--called Derjaguin approximation \cite{Derjaguin1934, Israelachvili2011}. All of these techniques provide only an approximation to the actual scaling functions: mean field theory is exact only for spatial dimension $\spdim\geqslant 4$ (with logarithmic corrections in $\spdim=4$); within the Monte Carlo simulations the size of the lattice is limited and therefore it is challenging to extract from the numerical data reliable results for the scaling limit; and within the Derjaguin approximation geometrical aspects of the system are not captured accurately. Out of these available methods, the Derjaguin approximation is the most straightforward scheme and requires the least numerical effort. Therefore, it provides an appropriate starting point for investigating the critical Casimir interaction in our system.

Within the Derjaguin approximation any available result for the planar geometry (typically the best one) can be used in order to estimate the effective interaction between more complicated objects. This approximation has been applied for many problems \cite{Israelachvili2011, Adamczyk1999, Thennadil2001, Oettel2004, Rentsch2006} and, in the case of the critical Casimir force, the results typically agree qualitatively (and under favorable circumstances even quantitatively) with the proper ones as far as they are available \cite{Hanke1998, Mohry2014, Labbe2016}. Here, we describe briefly the concept of this approximation (mostly in order to introduce those objects and quantities which turn out to be useful for our analysis); concerning the discussion of the validity of this approximation in our present case see Sec.~\ref{secE}.

We assume that the particles are in the special configuration (i.e.,~the center of the first sphere is at the origin and the center of the second one is on the positive $y$ semi--axis at the point $\left(r_x=0,r_y=2R+D,r_z=0\right)$). We introduce the \textit{projection plane}, parallel to the $x$ and $z$ axes, and project orthogonally onto it the patterns on the hemispheres of both colloids facing each other (i.e., the right hemisphere of the first particle and the left hemisphere of the second one). The resulting figure is a circle $\Lambda$ of radius $R$ which we separate into four disjoint regions (sets) $\Lambdapp$, $\Lambdapm$, $\Lambdamp$, and $\Lambdamm$. The first and second sign in the index of $\Lambda$ denotes the preference of the surface of the first and second particle, respectively. For example, for every point $P\in \Lambdapm$ on the projection plane there are two points, one on each sphere, which are projected onto $P$; the point on the first particle is in the patch while the point on the second particle is not. A complete example of the construction scheme described above is presented in Fig.~\ref{secC:pplane}. It is convenient to additionally define two sets $\Lambdasame$ and $\Lambdaopposite$, where the properties of the two points on the surfaces of both particles are the same and opposite, respectively:
\begin{equation}
\Lambdasame=\Lambdapp \cup \Lambdamm,\qquad \Lambdaopposite=\Lambdapm \cup \Lambdamp. 
\end{equation}
In order to make the definitions mathematically complete, it is necessary to define the surface affinity also at the edge of the patches. We assume that in these points the surface exhibits the same preference as the inside of the patch, i.e.,~the patch on the sphere is a closed set. As expected, our results do not depend on this convention.

Within the Derjaguin approximation, for each point $P$ on the projection plane, the distance $\ell$ between those two points which are projected onto $P$ is calculated, and the contribution to the force of the surface element $\dd A$ around the point $P$ is estimated via Eq.~\eqref{secC:cforceslab} to be
\begin{equation}\label{secC:dDerjaguinForce}
\dd \cforce=\frac{\kB \Tc}{\ell^3}\,\fscslab_\mathrm{s}\left(\omega\right)\dd A,
\end{equation}
where both the distance $\ell$ and the scaling variable $\omega$ depend on $P$, and `$\mathrm{s}$' denotes the pair of boundary conditions for the points projected onto $P$. The total force is obtained as the integral of Eq.~\eqref{secC:dDerjaguinForce} over the whole circle $\Lambda$ on the projection plane:
\begin{equation}\label{secC:DerjaguinGeneral}
\cforce=\int_{\Lambdasame} \!\frac{\kB \Tc}{\ell^3}\,\fscslab_\bsame\left(\omega\right)\dd A+\int_{\Lambdaopposite}\!  \frac{\kB \Tc}{\ell^3}\,\fscslab_\bopposite\left(\omega\right)\dd A.
\end{equation}

It is convenient to parametrize the circle $\Lambda$ on the projection plane by using the spherical coordinates of the first particle $0\leqslant\theta\leqslant\pi$ and $0\leqslant\phi\leqslant\pi$, where the range of $\phi$ is restricted because only the facing hemisphere of the first particle is used in the parametrization (see Fig.~\ref{secC:pplane}). For this choice of integral variables we have determined
\begin{subequations}\label{secC:parametrization}
\begin{align}
 \dd A&= R^2\sin^2\theta\,\sin\phi\,\dd \theta\,\dd \phi,\\
 \ell&=2 R\left(1-\sin\theta\,\sin\phi\right)+D,\\
\label{secC:parametrization_omega}  \omega &=\Theta\left[1+2\left(1-\sin\theta\,\sin\phi\right)/\Delta\right]\equiv \spomega,
\end{align}
\end{subequations}
where $R$ is the radius of the spherical colloids, $D$ is their surface--to--surface distance, and $\Delta$ and $\Theta$ are the scaling variables, given in Eq.~\eqref{secC:DeltaTheta}. In order to simplify the notation, we have denoted the expression in Eq.~\eqref{secC:parametrization_omega} by $\spomega$; it is an argument of the scaling functions for the slab geometry when they are used to calculate the scaling functions for two spheres within the Derjaguin approximation.

Using Eqs.~\eqref{secC:DerjaguinGeneral}, \eqref{secC:parametrization}, and \eqref{secC:forcesf}, we have calculated the formula for the scaling function for the critical Casimir force acting on the first particle:
\begin{multline} \label{secC:DerjaguinForce}
 \fscv^{\left(1\right)}\left(\Delta,\Theta, \sphconf^\ast\right)=\\
 \vctr{e}_y\int_0^\pi \dd \theta \int_0^\pi \dd \phi\, \frac{\Delta^2 \sin^2\theta\,\sin\phi}{\left[\Delta+2\left(1-\sin\theta\,\sin\phi\right)\right]^3}\\ 
 \times \fscslab_{\mathrm{s}\left(\theta,\phi\right)}\left(\spomega\right),
\end{multline}
where the variables $\theta$ and $\phi$ are the spherical angular coordinates on the first particle, $\mathrm{s}\left(\theta,\phi\right)$ denotes the same or opposite boundary conditions in the point parametrized by $\theta$ and $\phi$ ($\mathrm{s}$ depends on the configuration $\Omega^\ast$ of the particles), and the argument $\spomega$ of the scaling function is given by Eq.~\eqref{secC:parametrization_omega}.

The force as given by Eq.~\eqref{secC:DerjaguinForce} cannot be considered as a reliable approximation of the actual critical Casimir force acting between colloids with inhomogeneous surfaces. First, it always acts in the direction of the line connecting the centers of particles (i.e., it is a radial force). Second, within the Derjaguin approximation there is no straightforward way to determine the torques present in the system (besides the slab geometry in which there are none). Third, the interaction described by Eq.~\eqref{secC:DerjaguinForce} is not conservative (because a radial force is conservative if and only if it does not depend on the angles).

In order to overcome the above problems, we use the Derjaguin approximation for deriving the potential of interaction instead of deriving the force. Straightforward calculation leads to
\begin{multline}\label{secC:DerjaguinPotential}
 \psc\left(\Delta,\Theta,\Omega^\ast\right)=\\
 \int_0^\pi \dd \theta \int_0^\pi \dd \phi\, \frac{\Delta \sin^2\theta\,\sin\phi}{\left[\Delta+2\left(1-\sin\theta\,\sin\phi\right)\right]^2}\,\pscslab_{\mathrm{s}\left(\theta,\phi\right)}\left(\spomega\right), 
\end{multline}
where $\spomega$ is given by Eq.~\eqref{secC:parametrization_omega}. The critical Casimir forces and torques are calculated as derivatives of the potential. Accordingly, by construction the obtained interaction is conservative. The formulae for the scaling functions $\fscv^{\left(i\right)}$ and $\tscv^{\left(i\right)}$ are provided in Appendix~\ref{secV}.

It is reassuring that the Derjaguin approximation for the force (Eq.~\eqref{secC:DerjaguinForce}) renders the same expression for the radial component of the force as the corresponding derivative of the potential given in Eq.~\eqref{secC:DerjaguinPotential}. The difference between these two approaches is that in addition the potential gives the torques and the non--radial components of the force in such a way that the interaction is conservative.

\section{Method of calculation}\label{secD}

\subsection{Scaling functions for the slab geometry}\label{secD:sfslab}

In order to be able to calculate the scaling functions within the Derjaguin approximation, it is necessary to know the scaling functions for the slab geometry (see Sec.~\ref{secC:sg}). They can be estimated, e.g., via Monte Carlo simulations or, alternatively, by using the extended de~Gennes--Fisher local--functional method \cite{Borjan2008,Upton2013}.

\begin{figure}
\includegraphics[width=0.45\textwidth]{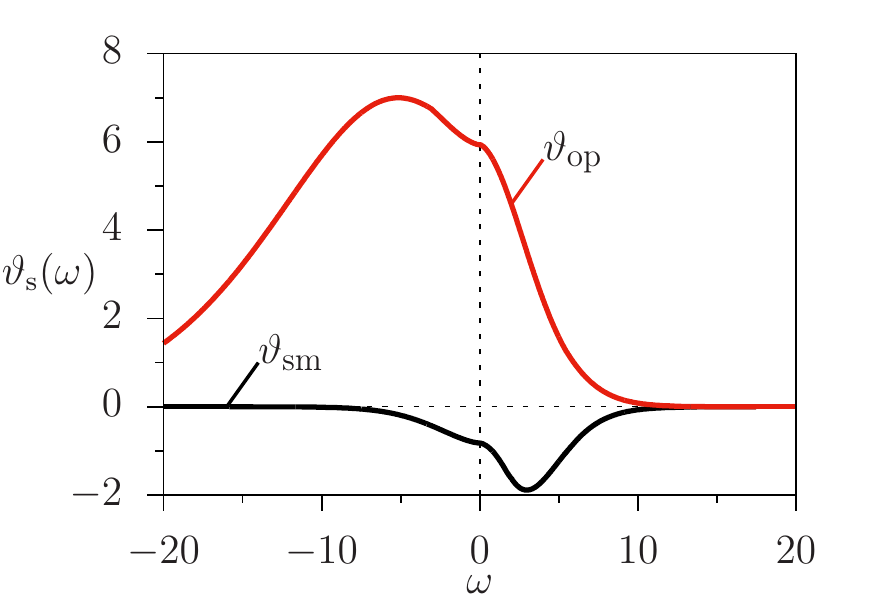} 
\caption{Scaling functions for the critical Casimir force, in the slab geometry (Eq.~(\ref{secC:cforceslab})) for the 3D Ising universality class, for opposite ($\fscslab_{\bopposite}$) and same ($\fscslab_{\bsame}$) boundary conditions as functions of the scaling variable $\omega$ (Eq.~(\ref{secC:scalingvariable})). For further details see Appendix~\ref{secX}.}
\label{secD:Slab-Scaling}
\end{figure}

Here we use the data from the corresponding Monte Carlo simulations \cite{Vasilyev2009a, *Vasilyev2009b}. This technique allows one to estimate the scaling functions only for a limited number of values of the scaling variable $\omega$. In order to obtain the full scaling functions $\fscslab_{\bsame}\left(\omega\right)$ and $\fscslab_{\bopposite}\left(\omega\right)$ it is necessary to interpolate and extrapolate the available data; the details of this procedure are described in Appendix~\ref{secX}. We plot the resulting scaling functions in Fig.~\ref{secD:Slab-Scaling}.

We note that the function $\fscslab_{\bsame}\left(\omega\right)$ is negative throughout (i.e., if both walls prefer the same component of the mixture, there is a critical Casimir attraction) whereas $\fscslab_{\bopposite}\left(\omega\right)$ is positive throughout (i.e., there is critical Casimir repulsion of walls preferring different liquid components). Additionally, the absolute value of the scaling function is larger in the case of opposing surface affinities.

Because of the unknown form of the leading corrections to scaling (see Ref.~\cite{Vasilyev2009a,*Vasilyev2009b}), the interpolation of the Monte Carlo data and thus, the construction of the above scaling functions suffer from numerical errors. The systematic error is estimated to be up to $20\%$ \cite{Labbe2016}, which translates directly to all of our numerical results. In order to increase the precision, one can normalize all results by dividing them by the critical Casimir amplitude $\fscslab_{\bsame}\left(0\right)$; such normalized functions have a numerical error of up to $5\%$. We note that in our calculations this systematic error, together with the inaccuracies of the Derjaguin approximation, is the main source of error; all other numerical inaccuracies present in our calculations are much smaller and can be neglected. 
On the other hand there is good reason to be confident about the reliability of the scaling functions shown in Fig.~\ref{secD:Slab-Scaling}, because they agree excellently with high resolution experimental data~\cite{Hertlein2008} available for $|\omega|\gtrsim1$.

\subsection{Calculation of the scaling function for the interaction potential}\label{secD:calcpot}

The scaling functions $\pscslab_\mathrm{s}$ (see Eq.~\eqref{secC:pscslabdef}) are prerequisites for calculating the critical Casimir potential of interaction for two spherical colloids (see Eq.~\eqref{secC:DerjaguinPotential}).

The calculation of the corresponding integral is based on the adaptive quadrature algorithm implemented in the GNU Scientific Library (GSL) \cite{Gough2009}. In the course of carrying out the integral in Eq.~\eqref{secC:DerjaguinPotential}, we first fix the value of $\theta$ and calculate the integral over $\phi$:
\begin{equation}
I_1\left(\theta\right)=\int_0^\pi I_0\left(\theta,\phi\right)\dd \phi,
\end{equation}
where $I_0\left(\theta,\phi\right)$ denotes the integrand in Eq.~\eqref{secC:DerjaguinPotential}. We note that the function $I_0$ is discontinuous at all points where the surface boundary conditions change. Moreover, for certain values of $\theta$ it can happen that two points of discontinuity are located very close to each other and the change of integrand is easy to miss in the numerical integration. In order to avoid this problem, for each $\theta$ we calculate analytically all those values of $\phi$, for which $I_0$ is discontinues and subdivide the integral as follows:
\begin{equation}\label{secD:splitone}
 I_1\left(\theta\right)=\int_0^{\phi_1}I_0 \dd\phi+\int_{\phi_1}^{\phi_2}I_0\dd\phi+\ldots+\int_{\phi_{k}}^{\pi}I_0\dd\phi,
\end{equation}
where $0<\phi_1<\phi_2<\ldots<\phi_k<\pi$ denotes all points where the integrand has a discontinuity. This way all discontinuities of $I_0$ are properly taken into account in the course of the integration. Additionally, as all integrands on the right--hand side of Eq.~\eqref{secD:splitone} are now continuous functions of $\phi$, this subdivision is reducing the time required for the numerical calculation.

Finally, we calculate the integral over $\theta$ in order to obtain the scaling function
\begin{equation}
\psc =\int_0^\pi I_1\left(\theta\right)\dd \theta.
\end{equation}
Here, we locate all values of $\theta$, at which the patches start or end, and split up the integral accordingly.

The resulting scaling function $\psc\left(\Delta,\Theta,\sphconf^\ast\right)$ is evaluated with a relative or an absolute error of $10^{-6}$, whichever is attained first.

The program performing the above algorithm of evaluation of the scaling function was written in C++. Further processing of the data was carried out using Mathematica~\cite{Mathematica2018}.

\subsection{Calculation of the forces and torques}\label{secD:numder}

In order to obtain the forces and torques acting on the colloids we use Eq.~\eqref{secV:forcetorque}. This implies that we have to calculate numerically the derivatives of the scaling function $\psc\left(\Delta, \Theta, \sphconf^\ast\right)$ for the potential. In this section we describe the corresponding procedure. 

First, we note that it is not necessary to calculate the derivatives $\partial \psc/\partial \Delta$ and $\partial \psc/\partial \Theta$. In the derivation of forces and torques both the radius $R$ of the particles and the temperature $T$ are fixed, and only the surface--to--surface distance $D$ can change (see Eq.~\eqref{secC:DeltaTheta}). Second, a close inspection of Eq.~\eqref{secV:forcetorque} shows that the derivative $\partial \psc/\partial D$ appears only in the formula for the radial component of the force (see Eq.~\eqref{secV:FIy}), which can be calculated directly from the Derjaguin approximation for the forces (see Eq.~\eqref{secC:DerjaguinForce}).

It remains to determine the derivatives of $\psc$ with respect to the angles appearing in the relative configuration $\Omega^\ast$. In order to simplify the notation, in the following we discuss the calculation of $\partial \psc/\partial \delta$, where $\delta$ is one of the angles $\beta_1^\ast$, $\gamma_1^\ast$, or $\beta_2^\ast$. The value of $\delta$, at which the derivative is calculated, is denoted as $\delta_0$. We also do not consider the dependence of $\psc$ on all other variables as they are fixed.

The general procedure of calculating $\left.\partial \psc/\partial \delta\right|_{\delta=\delta_0}$ is as follows: First, we fix a small positive number $\tmpvari$ and a positive integer $n$. Second, we calculate the values of the function $\psc$ for $n$ values of $\delta$ uniformly distributed in the interval $\mathcal{I}=\left[\delta_0-\tmpvari,\delta_0+\tmpvari\right]$. We denote these points in $\mathcal{I}$ by $\delta_i$ for $i=1,2,3,\ldots,n$. Third, we fit the quadratic function
\begin{equation}\label{secD:fitfunction}
 f\left(\delta\right)=a \delta^2+b \delta+c
\end{equation}
to the points $P_i=\left(\delta_i,\psc\left(\delta_i\right)\right)$ (obtained in the second step) by using the method of least squares. This way we obtain the values of the coefficients $a$, $b$, and $c$. With them the estimate of the derivative is
\begin{equation}
 \left.\frac{\partial \psc}{\partial \delta}\right|_{\delta=\delta_0}\approx f^\prime\left(\delta_0\right)=2 a \delta_0+b.
\end{equation}
The quadratic term in the fitting function in Eq.~\eqref{secD:fitfunction} has been included in order to somehow account for the fact that in general the function $\psc$ is not linear within the interval $\mathcal{I}$.

The method of least squares was used in order to reduce the error stemming from the chosen numerical integration scheme for calculating $\psc$; this error is similar to random noise. We note that other sources of error, like the inaccuracies of the scaling functions $\pscslab_{\bsame}$ and $\pscslab_{\bopposite}$ (see Eq.~\eqref{secC:pscslabdef} and Appendix~\ref{secX}), or inaccuracies of the Derjaguin approximation, are of a more systematic nature, and the fitting procedure leaves these errors unaltered.

The above general procedure needs to be adjusted near certain special values of $\delta_0$. One of these problems arises, if $\delta_0$ is located close to the boundary of the domain of the function $\psc$, and certain values $\psc\left(\delta_i\right)$ cannot be calculated. In this case, for the fitting we use only those points $\delta_i$ which are inside the domain. This way the number of points is reduced but it is still not smaller than $n/2$.

If there is a point of nonanalyticity of the function $\psc$ inside the interval $\mathcal{I}$, the fitting function in Eq.~\eqref{secD:fitfunction} might be not a good approximation. Around such a point, the general procedure fails and needs to be corrected. This is the reason why, in order to calculate the derivatives, it is necessary to investigate nonanalyticities of $\psc$. Below, in Sec.~\ref{secD:nonanalycities}, we present the observed types of singularities of the interaction and discuss how to adjust the general procedure described above.

If the point $\delta_0$ is far away from the points of nonanalyticity, the above procedure gives, as we have checked, reliable results for $\tmpvari=10^{-3}$ and $n=21$. These values have been used in order to calculate all results presented in Sec.~\ref{secK}.

\subsection{Nonanalyticities of scaling functions}\label{secD:nonanalycities}

\begin{figure*}
\begin{center}
 \includegraphics[width=0.9\textwidth]{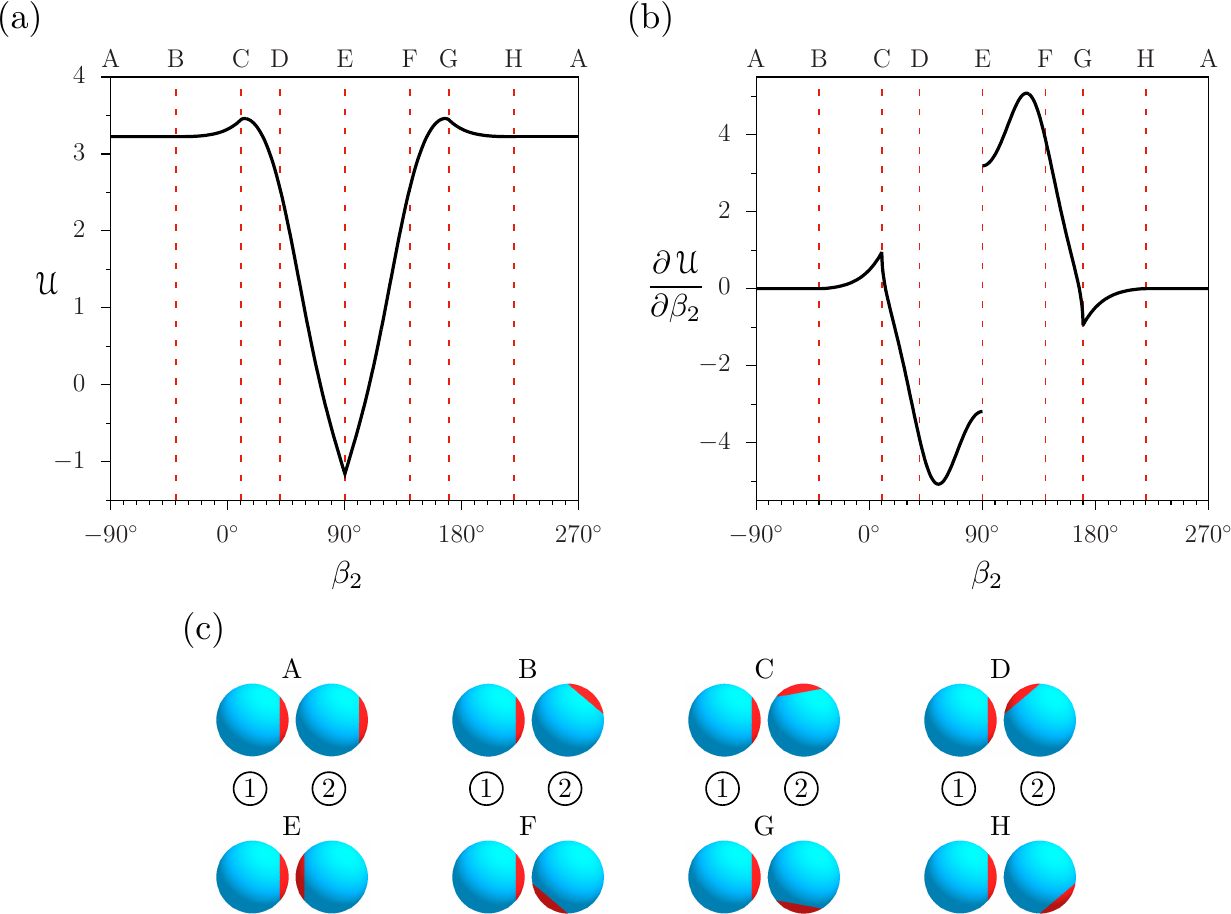}
\end{center}
\caption{\label{secD:fig1} Typical nonanalyticities of the scaling function for the potential $\psc\left(\Delta,\Theta,\Omega^\ast\right)$. (a) $\psc$ as a function of $\beta_2$ for $\Delta=\Theta=0.5$, $\beta_1=-90\degree$, and $\gamma_1=0$. (b) Derivative of $\psc$ with respect to $\beta_2$; the values of the other parameters are the same as in (a). Up to the factor $\Delta$, this quantity equals $\tsc_{x}^{\left(2\right)}$, which is the $x$ component of the scaling function for the torque acting on the second colloid. (c) The configurations of particles at special points. The letters A--H, denoting the configurations, are also marked at the top of the plots (a) and (b). In the configurations B--H the scaling function $\psc$ is not analytic. In Sec.~\ref{secD:nonanalycities} we classify these nonanalyticities into three types: type \singi singularity (configuration E in the plots) where there is a `V' shaped cusp nonanalyticity of $\psc$ and where there is a jump of the derivative of $\psc$; type \singii singularity (configurations C and G) for which $\psc$ exhibits a bump and for which the derivative of $\psc$ exists but has an infinite slope; and type \singiii singularity (configurations B, D, F, and H) where the second derivative of $\psc$ has an infinite slope. See the main text for further details.}
\end{figure*}

Upon changing the relative orientation $\sphconf^\ast$ of the colloids, we have observed three main types of singularities in the interaction of the colloids. In this subsection, we briefly characterize them and describe how the derivatives of the scaling functions around them can be calculated numerically. A more detailed mathematical analysis is presented in Appendix~\ref{secW}.

For reasons of simplicity, we consider the particles in a special configuration (see Sec.~\ref{secB}). In Fig.~\ref{secD:fig1}(a) we present the typical behavior of the scaling function for the potential $\psc$ when one of the particles is rotated around the $x$ axis; it exemplifies the three main types of singularities. In Fig.~\ref{secD:fig1}(b) we plot the derivative of the scaling function for the interaction potential, which is actually proportional to the $x$ component $\tsc_{x}^{(2)}$ of the vectorial scaling function for the torque (see 
Eq.~\eqref{secV:TIIx}). We note that in this figure we allow for $\beta_2>90\degree$ --- in this case the rotation is equivalent to the one with $\tilde{\beta}_2=180\degree-\beta_2$ and $\tilde{\gamma}_2=180\degree$. (Here the superscript `${}^\ast$' can be omitted because the initial configuration is already a special one.)

The singularity of type \singi (type one) appears if the patches on both colloids form a so--called \textit{mirror--symmetric configuration}, i.e., for $\beta_2=-\beta_1>-\thp$ and $\gamma_1=0$, like configuration E in Fig.~\ref{secD:fig1}(c). In this case, $\Lambdasame=\Lambda$ and (within the Derjaguin approximation) the function $\psc$ has the same value as for homogeneous spheres. If in this case any of the colloids is rotated, $\Lambdaopposite$ becomes a nonempty set and $\psc$ increases. This produces a characteristic `V' shaped cusp of the scaling function for the potential and a discontinuity of its first derivative (see Fig.~\ref{secD:fig1} for $\beta_2=90\degree$). In Appendix~\ref{secW} we show that the left-- and the right--side derivatives of the function $\psc$ at a point of type \singi singularity have the same absolute value but opposite signs.

In order to calculate numerically the derivative of $\psc$ close to the nonanalyticity of type \singi, the procedure described in Sec.~\ref{secD:numder} has to be modified. For the fitting, instead of Eq.~\eqref{secD:fitfunction}, we use
\begin{equation}
f_{\text{\singi}}\left(\delta\right)=a_{\singi}\left(\delta-\delta_{\singi}\right)^2+b_{\singi}\left|\delta-\delta_{\singi}\right|+c_{\singi},
\end{equation}
where $\delta_{\singi}$ denotes the value of the angle, for which the singularity occurs; $a_{\singi}$, $b_{\singi}$, and $c_{\singi}$ are fitting parameters. Exactly at $\delta_{\singi}$ the derivative does not exist. However, because at $\delta_{\singi}$ the potential reaches its minimum value, one can assume that the derivative at $\delta_{\singi}$ is equal to zero, i.e., in the configuration with patches in mirror--symmetric configuration, there are no non--radial forces and torques.

The occurrence of singularities of type \singi in the scaling function of the critical Casimir interaction, which is calculated within the Derjaguin approximation for Janus particles, has been reported in Ref.~\cite{Labbe2016}.

The singularity of type \singii occurs if the projections of the two patches on the projection plane are tangent, i.e., if there is only a single point in the region $\Lambdapp$; see the configurations C and G in Fig.~\ref{secD:fig1}(c). In order to analyze this case it is useful to introduce the overlap angle
\begin{equation}\label{secD:overlapb}
 \zeta_\text{overlap}^{\singii}=2\thp-\zeta_\mathrm{pp},
\end{equation}
where $\zeta_\mathrm{pp}$ is the angular distance between the midpoints of the patches, i.e., the angle between $\vctr{n}_1$ and $\vctr{n}_2$, where $\vctr{n}_i$, for $i=1,2$, is a unit vector starting from the center of the $i$--th particle and ending at the surface of the same particle, in the midpoint of the patch.

If $\zeta_\text{overlap}^{\singii}>0$, the patches overlap and the region $\Lambdapp$ is nonempty. When $\zeta_\text{overlap}^{\singii}<0$ there is no overlap and, on the projection plane, the two regions $\Lambdapm$ and $\Lambdamp$ are separated by $\Lambdamm$. The nonanalyticity of $\psc$ occurs for $\zeta_\text{overlap}^{\singii}=0$. In Appendix~\ref{secW} we show that the nonanalytic part of $\psc$ is proportional to
\begin{equation}
 \psc^{\singii}_{\text{nonanalytic}}\propto\begin{cases}
  \left(\zeta_\text{overlap}^{\singii}\right)^{3/2}, & \zeta_\text{overlap}^{\singii}\geqslant 0,\\
  0, & \zeta_\text{overlap}^{\singii}<0,
 \end{cases}
\end{equation}
with the corrections of the order of $\left(\zeta_\text{overlap}^{\singii}\right)^{5/2}$. This behavior makes the nonanalycities of type \singii hard to notice in the plots of the scaling function for the potential (especially if the amplitude of the nonanalytic term is small). On the other hand, the derivative of $\psc$ behaves like a square root (i.e., a cusp with infinite slope) and thus in the plots of the forces and of the torques these singularities are well visible; see Fig.~\ref{secD:fig1} for $\beta_2=10 \degree$ and $170\degree$ (configurations C and G).

In order to calculate the derivative of $\psc$ with respect to a certain angle $\delta$ in the region close to a singularity of type \singii, the procedure described in Sec.~\ref{secD:numder} has to be modified by replacing the fitting function in Eq.~\eqref{secD:fitfunction} with
\begin{equation}\label{secD:fbump}
 f_{\singii}\left(\delta\right)=a_{\singii} \delta^2+b_{\singii} \delta +c_{\singii}+d_\singii \,\Heaviside\left[\varkappa\left(\delta-\delta_{\singii}\right)\right] \left|\delta-\delta_{\singii}\right|^{3/2}, 
\end{equation}
where $a_{\singii}$, $b_{\singii}$, $c_{\singii}$, and $d_{\singii}$ are fitting parameters; $\delta_{\singii}$ is the value of $\delta$ for which there is a nonanalyticity; $\Heaviside$ denotes the Heaviside step function; and $\varkappa=+1$ if $\zeta_\text{overlap}^{\singii}$ increases upon increasing $\delta$, and $\varkappa=-1$ otherwise. The additional term accounts for the nonanalytic part of $\psc$. We note that, unlike for the singularity of type \singi, at $\delta_0=\delta_{\singii}$ the first derivative of $\psc$ exists but the second derivative diverges from one side. Up to our knowledge, this is the first report of this type of nonanalyticity for the critical Casimir interaction calculated within the Derjaguin approximation.

Within the Derjaguin approximation, only half of each sphere takes part in the interaction, i.e., points on the right hemisphere of the first colloid interact with points on the left hemisphere of the second colloid. If the patch is fully located within one of the two other hemispheres, the energy of interaction does not depend on its precise position and the derivatives with respect to some of the angles in $\sphconf^\ast$ are zero. Accordingly, a nonanalyticity of the function $\psc$ is expected to occur if the patch passes through the great circle separating the two hemispheres. We call this type of singularity of the interaction as type \singiii. It is expected to occur if the edge of the patch is tangent to the great circle separating the interacting and noninteracting hemispheres, i.e., if on the projection plane one of the sets $\Lambdapp \cup \Lambdapm$ or $\Lambdapp \cup \Lambdamp$ have exactly one point in common with the edge of the circle $\Lambda$; this situation takes place if $\left|\beta_1\right|=\thp$ or $\left|\beta_2\right|=\thp$. This nonanalyticity occurs for the configurations B, D, F, and H in Fig.~\ref{secD:fig1}(c).

Also in this case it is useful to define the two overlap angles
\begin{equation}\label{secD:overlapSR}
 \zeta_{\text{overlap},i}^{\singiii}=90\degree+\thp-\zeta_{\mathrm{ph},i}, \quad i=1,2,
\end{equation}
where the patch--hemisphere angle $\zeta_{\mathrm{ph},1}$ ($\zeta_{\mathrm{ph},2}$) denotes the angle between the unit vector $\vctr{n}_1$ ($\vctr{n}_2$) pointing to the midpoint of the patch on the first (second) colloid (see the discussion after Eq.~\eqref{secD:overlapb}), and the vector $\vctr{n}_{\mathrm{h},2}=\vctr{e}_y$ ($\vctr{n}_{\mathrm{h},1}=-\vctr{e}_y$) pointing to the midpoint of the noninteracting hemisphere on the second (first) colloid. The singularity of type \singiii occurs in configurations for which $\zeta_{\text{overlap},i}^{\singiii}=0$ or $\zeta_{\text{overlap},i}^{\singiii}=2\thp$ for $i=1$ or $i=2$.

In Appendix~\ref{secW} we argue that singularities of type \singiii manifest themselves via a nonanalytic term in $\psc$ of the form
\begin{subequations}
\begin{align}
\psc^{\singiii}_{\text{nonanalytic}}\propto&
\begin{cases}
 \left(\zeta_{\text{overlap},i}^{\singiii}\right)^{5/2}, & \zeta_{\text{overlap},i}^{\singiii}\geqslant 0,\\
 0, &\zeta_{\text{overlap},i}^{\singiii}<0,
\end{cases}\\
\psc^{\singiii}_{\text{nonanalytic}}\propto&
\begin{cases}
 \left(2\thp-\zeta_{\text{overlap},i}^{\singiii}\right)^{5/2}, & \zeta_{\text{overlap},i}^{\singiii}\leqslant 2\thp,\\
 0, &\zeta_{\text{overlap},i}^{\singiii}>2\thp,
\end{cases}
\end{align}
\end{subequations}
where the first term is relevant for $|\zeta_{\text{overlap},i}^{\singiii}|\ll 1$, while the second term holds for $|\zeta_{\text{overlap},i}^{\singiii}-2\thp|\ll 1$. This means that the derivatives of $\psc$ with respect to the angles (i.e., forces and torques) exhibit a nonanalyticity of the order of $\delta^{3/2}$.

For $|\zeta_{\text{overlap},i}^{\singiii}|\ll 1$ the nonanalytic term in the formulae for the force and torque is the leading order term and therefore it is well visible. In contrast, for $\left|\zeta_{\text{overlap},i}^{\singiii}- 2\thp\right|\ll 1$ the nonanalytic term is only a small correction and other analytic terms dominate. In the latter case, within our accessible numerical precision, we have not been able to verify that the function $\psc$ contains this term.

Since the singularity of type \singiii for the potential manifests itself through a nonanalyticity of the order of $5/2$ --- which is higher than the order of the polynomial in Eq.~\eqref{secD:fitfunction} used for the fitting (see Sec.~\ref{secD:numder}) --- there is no need to adjust the general procedure in this case.

The singularities discussed above are relevant for calculating the derivatives with respect to angles and, via Eq.~\eqref{secV:forcetorque}, for all components of the scaling functions for the force and torque, except for the radial components of the force $\fsc_{y}^{\left(i\right)}$. The resulting nonanalyticities of $\fscv^{\left(i\right)}$ and $\tscv^{\left(i\right)}$ can take three forms: jumps of the value, $\propto\delta^{1/2}$, and $\propto\delta^{3/2}$ for the singularities of type \singi, \singii, and \singiii, respectively (see Fig.~\ref{secD:fig1}(b)). 

In the case of the radial components of the scaling functions for the force, the situation is different. They are calculated from Eq.~\eqref{secV:FIy}, where there are no derivatives with respect to the angles. Therefore we expect that for these components the singularities of interaction should manifest themselves in the same way as they do for the scaling function $\psc$ for the potential --- the nonanalyticities of the form of a `V' shaped cusp, $\propto\delta^{3/2}$, and $\propto\delta^{5/2}$ for the singularities of type \singi, \singii, and \singiii, respectively. In practice, these components are calculated directly from Eq.~\eqref{secC:DerjaguinForce}. The properties of the radial component of the scaling function for the force are presented in Sec.~\ref{secK:radial}.

Finally we emphasize that all three types of singularities discussed above follow from using the Derjaguin approximation in its present form; the scaling functions for the force and the torque onto objects of finite size are expected to be analytic functions of their rotational configuration and scaled distance. Nonetheless, it is useful to have an overview of the properties of the present, widely used, approximation scheme.

\section{Validity of the Derjaguin approximation}\label{secE}

Before presenting our numerical results it is necessary to discuss the reliability of the Derjaguin approximation as formulated in Sec.~\ref{secC:Derjaguin}. 
Unfortunately, there is no systematic way to estimate the inherent error of this approximation. Moreover, up to our knowledge, so far the patchy particles considered here have not been studied by different techniques. Therefore we cannot estimate the accuracy of our results by making a simple comparison with other independent results. Instead, we look at similar models in order to identify possible shortcomings of the Derjaguin approximation.


We start with the case of homogeneous spherical particles of radius $R$ separated by a surface--to--surface distance $D$. At $\Tc$ for such a system, the potential following from using the Derjaguin approximation diverges $\propto D^{-1}$ for $D\to 0$ and vanishes exponentially for $D\to\infty$. General arguments based on conformal invariance \cite{Burkhardt1995, *Burkhardt1997} predict that for homogeneous spheres the critical Casimir potential diverges $\propto D^{-1}$ for small $D$, and decays $\propto D^{-\beta/\nu}$ for large $D$. This means that in this case the Derjaguin approximation fails to predict the long--ranged, algebraic decay. The detailed analysis shows that the Derjaguin approximation provides reliable results if $D\ll R$ \cite{Hanke1998, Schlesener2003, Hasenbusch2013}. In the scaling limit, the approximation is exact for $\Delta=D/R\to 0$ \cite{Hanke1998, Hasenbusch2013} but it fails to reproduce the correct behavior for large $\Delta$.

If the surfaces of the immersed objects are not homogeneous, the situation is different in that the Derjaguin approximation can give wrong results even for $\Delta\to0$. So far, results are available for inhomogeneous wall--wall \cite{Sprenger2006, Nowakowski2016}, sphere--wall \cite{Troendle2009}, and cylinder--wall \cite{Labbe2014, Labbe2016} geometries. Within mean field theory they are based on the numerical minimization of the corresponding Hamiltonian, which gives correct results (up to logarithmic corrections) in $\spdim=4$ spatial dimensions. Beside that, the approximation can be tested against Monte Carlo simulations, an exact solution in $\spdim=2$ \cite{Nowakowski2016}, and, to some extent, experimental results in $\spdim=3$ \cite{Troendle2009}.

The analysis of available data allows us to identify two general situations in which the Derjaguin approximation for the critical Casimir force is expected to give wrong results: (i) The characteristic length--scale of the pattern on the surface is much smaller than either the correlation length or the distance between walls. (ii) There is a region between interacting objects, where the gradient of the order parameter $\Phi\left(\vctr{r}\right)$ has a large component in the direction perpendicular to the direction $\vctr{n}_\mathrm{D}$ used for the projections as they are applied within the Derjaguin approximation (in our case $\vctr{n}_\mathrm{D}=\vctr{e}_y$), i.e., in regions with
\begin{equation}\label{secE:criterion}
\left|\bm{\nabla} \Phi\right|\gg \left|\bm{n}_\mathrm{D}\cdot \bm{\nabla}\Phi\right|. 
\end{equation}
Since obtaining the order parameter profile $\Phi\left(\vctr{r}\right)$ is typically at least as difficult as calculating the critical Casimir force, the above criterion is not useful for an exact analysis. Nevertheless, it is usually possible to roughly estimate the order parameter profile for a given system and then use Eq.~\eqref{secE:criterion} in order to check the validity of the Derjaguin approximation.

In situation (i), typically, the pattern on the surface is averaged and yields a certain effective homogeneous surface~\cite{Troendle2015}; this aspect is not captured correctly by the Derjaguin approximation. Since in our present study we consider only colloids with a single patch, here this case is not relevant unless the patch is very small ($\thp\ll 1$).

In situation (ii), the free energy associated with the rapid change of the order parameter in all but one direction is completely missed by the approximation. The most prominent example of such a case is the capillary bridge formation at $T<\Tc$ between patches on the two spheres. In this case, the force in the normal direction is increased by the contribution stemming from the surface tension of the interface present in the system \cite{Bauer2000, Willett2000, Malijevsky2015}. This effect is not captured within the Derjaguin approximation. For $T>\Tc$ the situation (ii) can occur if $D$ is small, i.e., in the region where the Derjaguin approximation is actually expected to work quite well.

We note that the above criteria are only conjectures based on a limited amount of data available in the literature, and they do not provide a strict answer to which extent the Derjaguin approximation is reliable or not; they are supposed to identify regions, where the approximation can fail. This issue definitely deserves more research. An example for the case in which, despite of the large perpendicular gradients of the order parameter, the approximation can provide quite accurate results, concerns the lateral component of the force in a system with a capillary bridge. It can be correctly described within the Derjaguin approximation, except in the vicinity of the breaking transition~\cite{Nowakowski2016}.

It is also important to mention, that the nonanalyticities of the critical Casimir force reported above in Sec.~\ref{secD:nonanalycities} are all related to the discontinuous variation of the chemical surface properties. Such changes are known to not propagate in this form into the fluctuating medium \cite{Sprenger2005}. Moreover, such nonanalyticities, up to our knowledge, have not been reported in calculations which are not based on the Derjaguin approximation. Therefore, we strongly expect that these singularities are an artifact of the Derjaguin approximation.

Finally, we note that both Monte Carlo simulations and mean field calculations for systems with two inhomogeneous spheres are numerically challenging. Up to our knowledge, such studies have not yet been reported, and most probably will give estimates of the scaling functions only for a rather limited number of points. The present results can provide guidance for how to interpolate these data points correctly, even in regions where the Derjaguin approximation is not working well.

\section{Results}\label{secK}

In this section we present our results for the critical Casimir interaction between patchy particles. Using the method described in Sec.~\ref{secD}, we have been able to calculate, within the Derjaguin approximation, the potential and all components of the forces and torques arising from the critical Casimir interaction between two spherical colloids with chemically inhomogeneous surfaces. In the special case of Janus particles ($\thp=90\degree$, i.e., the patch is covering half of the particle surface) some results have already been reported \cite{Labbe2016}; our analysis is in full agreement with them. Moreover, we have been able to extend these results by providing non--radial components of the critical Casimir force as well as of the critical Casimir torque.

In this section, we first discuss our results for the radial critical Casimir force and the potential, and compare them with those for the special case of Janus particles. Then, we present our results for the non--radial components of the critical Casimir force and for the torque. Finally, we present a comparison with corresponding experimental data.

\subsection{Radial component of critical Casimir force}
\label{secK:radial}

We start the discussion of our results by analyzing $\fsc_r^{\left(i\right)}$ for $i=1,2$, i.e., the radial component of the vectorial scaling function of the critical Casimir force acting on the first and second particle, respectively. Since $\fsc_r^{\left(1\right)}=-\fsc_r^{\left(2\right)}$ (see Eq.~\eqref{secC:Frelation}), we can focus on the force acting on the second particle. Note that in a special configuration, (i.e., the second colloid is located on the $y$ axis) the radial components are equal to $\fsc_y^{\left(i\right)}$.

\begin{figure}
\includegraphics[width=0.45\textwidth]{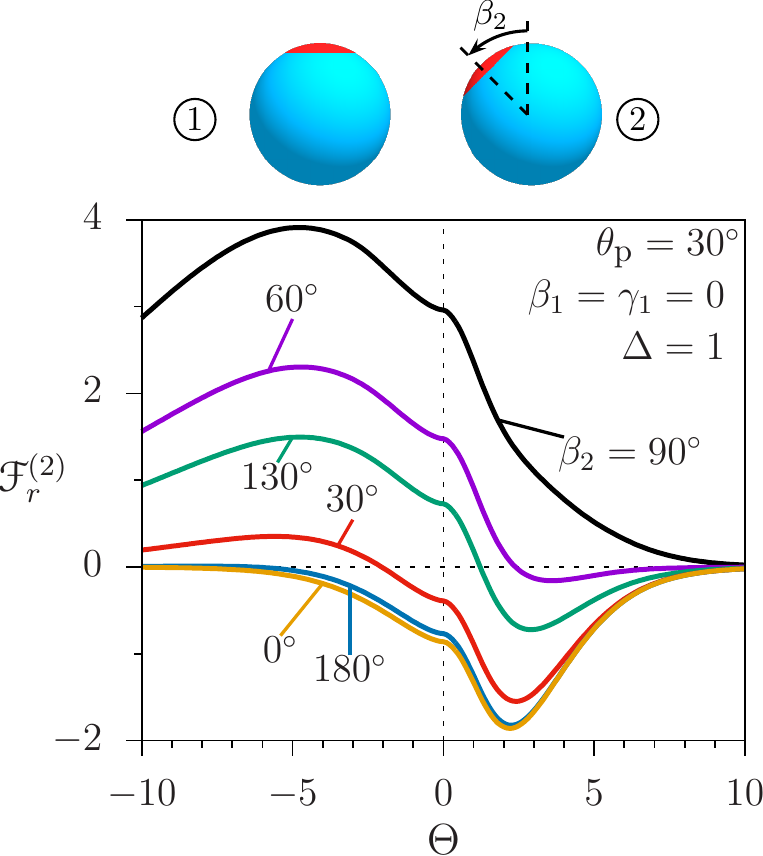} 
\caption{Scaling function $\fsc_r^{\left(2\right)}$ of the radial component of the critical Casimir force between two spherical colloids as function of $\Theta=\left(D \sign t\right)/\xib$ with $\xib=\xi_0^\pm\left|t\right|^{-\nu}$ and for a fixed value of $\Delta=D/R=1$ and various values of the rotation angle $\beta_2$. Positive values of $\fsc_r^{\left(2\right)}$ correspond to repulsion of colloids while negative ones correspond to attraction. A schematic plot of the configuration of the particles is presented above the graph.}
\label{secK:Fsc_r_vs_Theta}
\end{figure}

The dependence of $\fsc_r^{\left(2\right)}$ on the scaling variable $\Theta=\left(D\sign t\right)/\xib$ with $\xib=\xi_0^\pm \left| t\right|^{-\nu}$ for various values of the angle $\beta_2$ is presented in Fig.~\ref{secK:Fsc_r_vs_Theta}. For all plotted curves we have chosen $\Delta=D/R=1$, $\beta_1=\gamma_1=0$, and $\thp=30\degree$. If $\beta_2=0$ (i.e., if the patch on the second particle is in the topmost position), the radial component is negative (i.e., attractive) for all values of $\Theta$. Upon increasing $\beta_2$ (i.e., rotating the second particle such that its patch is moved towards the first particle), for any fixed $\Theta$ the radial component changes sign. This change occurs first for large negative values of $\Theta$. If $\beta_2=90\degree$ (i.e., the patch is facing the first particle), $\fsc_r^{\left(2\right)}$ is fully positive (i.e., repulsive). Upon increasing $\beta_2$ further, the strength of the repulsion decreases and, eventually, attraction is recovered. This change starts at large positive values of $\Theta$ and moves towards negative values of $\Theta$.

The observed behavior of the radial component of the scaling function of the critical Casimir force can be easily understood. If $\beta_2=0$, within the Derjaguin approximation, only the `$++$' and the `$--$' boundary conditions are active, so that due to $\fscslab_{\bsame}\left(\omega\right)<0$ the resulting net radial component is negative. If $\beta_2$ is increased, the region $\Lambdamp$ on the projection plane (see Sec.~\ref{secC:Derjaguin}) emerges (i.e., the set $\Lambdamp$ becomes nonempty) and, because $\fscslab_{\bopposite}\left(\omega\right)>0$, the force becomes less negative; if the patch is sufficiently large, the sign of the force eventually changes. The area of the region $\Lambdapp$ decreases upon increasing $\beta_2$, and $\Lambdapp$ becomes empty for $\beta_2>2\thp$. Starting from there, the dependence of the radial component of the scaling function on $\beta_2$ is solely determined by the location of the patch on the second particle; therefore we focus on the region $\Lambdamp$. For $\beta_2=90\degree$, the mean distance $\prdist$ between the points on the two spheres projected onto the region $\Lambdamp$ (see Sec.~\ref{secC:Derjaguin}) is smallest, and, moreover, the area of $\Lambdamp$ is largest; thus the repulsive contribution is strongest. Increasing $\beta_2$ further moves the patch to the bottom, where the mean distance $\prdist$ is large and, concomitantly, the area of $\Lambdamp$ shrinks. This explains the recovery of the attraction in this regime. Finally, we note that if the size of the patch $\thp$ is not large enough, the repulsive effect may not be sufficiently strong in order to change the sign of the radial component of the scaling function.

\begin{figure}
\includegraphics[width=0.45\textwidth]{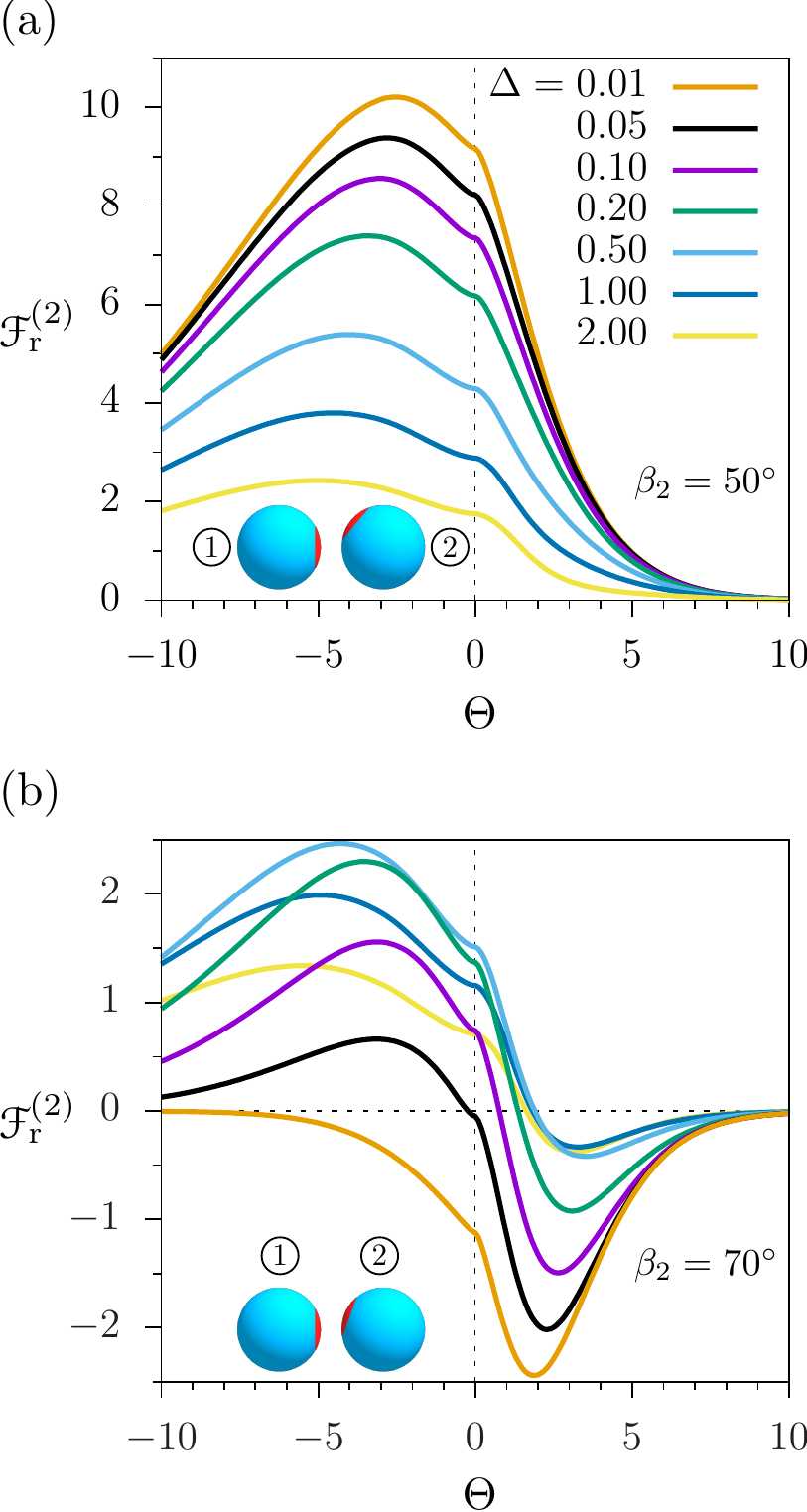} 
\caption{Radial component $\fsc_r^{\left(2\right)}$ of the scaling function for the critical Casimir force acting on the second particle as a function of $\Theta$ for various values of $\Delta$ and for two fixed orientations of both particles: $\thp=30\degree$, $\beta_1=-90\degree$, $\gamma_1=0$, and (a) $\beta_2=50\degree$ and (b) $\beta_2=70\degree$. Schematic plots of the particles are presented in each graph.}
\label{secK:Fsc_r_vs_Theta_Delta}
\end{figure}

Figure \ref{secK:Fsc_r_vs_Theta_Delta} shows the dependence of $\fsc_r^{\left(2\right)}$ on the scaling variable $\Delta$. Therein, both plots correspond to $\thp=30\degree$, $\beta_1=-90\degree$, and $\gamma_1=0$. For $\beta_2=50 \degree$ (see Fig.~\ref{secK:Fsc_r_vs_Theta_Delta}(a)) the radial component of the scaling function is always positive (i.e., there is repulsion) and has a maximum for $\Theta<0$. Upon increasing the scaled distance $\Delta$ between the particles, the strength of the interaction decreases together with a shift of the position of the maximum towards more negative values of $\Theta$.

The situation is quite different for $\beta_2=70\degree$ (see Fig.~\ref{secK:Fsc_r_vs_Theta_Delta}(b)). In this case, for $\Delta=0$, the function $\fsc_r^{\left(2\right)}$ is negative (corresponding to attraction) for all values of $\Theta$. Upon increasing the scaled distance $\Delta=D/R$ between the particles, the value of the scaling function grows for all values of $\Theta$. This leads to a change of sign of $\fsc_r^{\left(2\right)}$ for $\Theta<0$; this change sets in for large negative values of $\Theta$ and, upon increasing $\Delta$, it propagates towards larger values of $\Theta$. Starting from $\Delta\approx0.05$ the particles repel each other for all $\Theta<0$ (i.e.,~in the demixed region of the binary solvent), and, upon a further increase of $\Delta$, we observe the repulsion even for small positive values of $\Theta$. This behavior continues until $\Delta\approx 0.5$ is reached. Upon further increase of the scaled distance between the particles, the magnitude of the function $\fsc_r^{\left(2\right)}$ starts to decay. For $\Delta\neq0$ and $\Theta<0$ we observe a maximum of the radial component. If $\Delta$ is very large or very small the maximum is very broad and located at a very large negative value of $\Theta$. For $\Theta>0$ (i.e.,~in the mixed region of the binary liquid mixture) the radial component of the scaling function has a minimum. Upon increasing $\Delta$ the minimum becomes less deep and moves towards higher values of $\Theta$. This behavior of the minimum is slightly altered for $\Delta\approx 2$, where, upon increasing $\Delta$, the minimum moves towards smaller values of $\Theta$ and becomes deeper. This anomaly appears in the region where the Derjaguin approximation is expected to be unreliable so that it is physically irrelevant; we refrain from a more detailed discussion of this phenomenon.

The behavior described above can be understood by referring to the properties of the critical Casimir force in the slab geometry. If the scaled distance $\Delta$ between the particles is very small, the region where the surfaces are closest to each other contributes the most to the mutual interaction (due to the prefactor $L^{-3}$ in Eq.~\eqref{secC:cforceslab}). For $\beta_2=50\degree$ the boundary condition at the point of smallest distance $\prdist$ is `$+-$'; thus for small $\Delta$ the particles repel each other. On the other hand, if $\beta_2=70\degree$, the boundary condition at the same point is `$++$' and the particles attract each other. If $\Delta$ is increased, regions with larger values of $\prdist$ become relevant. In the case of $\beta_2=50\degree$, the attractive contribution stemming from the region $\Lambdasame$ is not sufficiently strong to dominate the radial component of the critical Casimir force. This is not surprising because the magnitude of $\fscslab_{\bopposite}\left(\omega\right)$ is larger than the magnitude of $\fscslab_{\bsame}\left(\omega\right)$ (see Fig.~\ref{secD:Slab-Scaling}). Additionally, below the critical point, $\fscslab_{\bsame}\left(\omega\right)$ is very small while $\fscslab_{\bopposite}\left(\omega\right)$ has a maximum. This is the reason why for $\Theta<0$ the value of $\fsc_r^{\left(2\right)}$ grows rapidly upon increasing $\Delta$. Above $\Tc$ the absolute value of the function $\fscslab_{\bsame}$ has a maximum, while $\fscslab_{\bopposite}$ is relatively small. This is the reason why, upon increasing $\Delta$, the radial component of the scaling function does not change sign for large values of $\Theta$.

The characteristic narrow plateau of the scaling function $\fsc_r^{\left(i\right)}$ around $\Theta=0$, which is visible for all curves in Figs.~\ref{secK:Fsc_r_vs_Theta}~and~\ref{secK:Fsc_r_vs_Theta_Delta}, is inherited from the scaling functions $\fscslab_{\bsame}\left(\omega\right)$ and $\fscslab_{\bopposite}\left(\omega\right)$ for the slab geometry. For small $\left|\omega\right|$ these functions behave like $A_1+A_2\left|\omega\right|^{1/\nu}$~\cite{Krech1994}, where $A_1$ and $A_2$ are constants.

\begin{figure}
\includegraphics[width=0.45\textwidth]{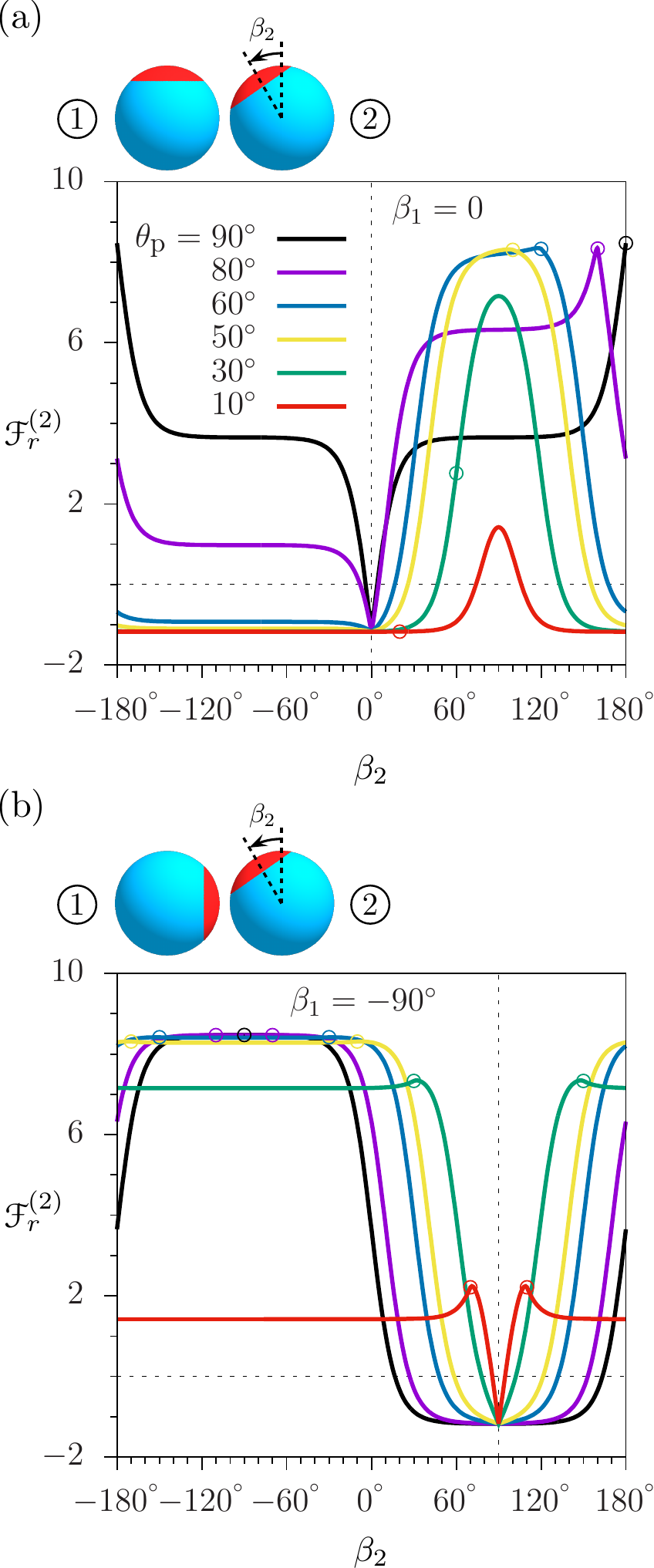} 
\caption{Scaling function $\fsc_r^{(2)}$ of the radial component of critical Casimir force, acting on the second particle, as function of $\beta_2$ for $\Theta=0$, $\Delta=0.2$, and $\gamma_1=0$, and for various sizes $\thp$ of the patch. The configuration of the first particle corresponds to (a) $\beta_1=0$ and (b) $\beta_1=-90\degree$. The circles indicate the singularities $\propto\left|\beta_2-\beta_{2,\text{sing.}}^{\left(i\right)}\right|^{3/2}$ of type \singii in $\fsc_r^{(2)}$ at various positions $\beta_{2,\text{sing.}}^{\left(i\right)}$. The color code in both panels is the same.}
\label{secK:compare1}
\end{figure}

The influence of the patch size $\thp$ on $\fsc_r^{(2)}$ is presented in Fig.~\ref{secK:compare1}. In this figure, $\fsc_r^{(2)}$ is plotted as a function of $\beta_2$ for various values of $\thp$, for fixed $\Theta=0$, $\Delta=0.2$, and $\gamma_1=0$, and for two orientations of the first particle: $\beta_1=0$ (Fig.~\ref{secK:compare1}(a)) and $\beta_1=-90\degree$ (Fig.~\ref{secK:compare1}(b)).

In Fig.~\ref{secK:compare1}(a), for $\thp\leqslant 45\degree$ the maximum of the radial component of the scaling function of the critical Casimir force is exactly at $\beta_2=90\degree$, in which case the center of the patch on the second particle is in the closest possible position to the first particle. In this position the area of $\Lambdamp$ is largest. If $\thp>45\degree$, for $\beta_2=90\degree$ the region $\Lambdapp$ becomes a nonempty set and, as a result, the radial force decreases because for elements of $\Lambdapp$ one has a negative contribution in Eq.~\eqref{secC:DerjaguinForce}. This effect shifts the position of the maximum towards higher values of $\beta_2$. Upon increasing $\thp$, starting from $45\degree$, the position of the maximum increases towards $180\degree$, which is attained for $\thp=90\degree$. Upon this increase the maximum becomes sharper. In Fig.~\ref{secK:compare1}(a), for $\thp<90\degree$, we observe a singularity $\propto \left(\beta_2-\beta_{2,\text{sing.}}^{\left(i\right)}\right)^{3/2}$ of type \singii located at the values $\beta_{2,\text{sing.}}^{\left(i\right)}$ of $\beta_2$ for which the patches are tangent (these points are marked by circles in the plot). These singularities are located to the left of the maximum for $\thp<45\degree$, coincide with the maximum for $\thp=45\degree$ and $\thp=90\degree$, and for $45\degree<\thp<90\degree$ they are shifted slightly to the right side of the maximum. We note that for $\thp=90\degree$ the radial force has an inverse `V' shape around $\beta_2=180\degree$.

If $\beta_1=-90\degree$, as shown in Fig.~\ref{secK:compare1}(b), for any size of the patch $\thp$ the radial component of the scaling function is symmetric around $\beta_2=90\degree$ (and around $\beta_2=-90\degree$). For $\beta_2=90\degree$ the patches are facing each other and (within the Derjaguin approximation) $\fsc_r^{\left(2\right)}$ does not depend on the size $\thp$ of the patch. Upon decreasing $\beta_2$, a region $\Lambdaopposite$ with opposing boundary conditions emerges and thus contributing to repulsion so that the radial component grows. Since the magnitude of $\fscslab_{\bopposite}$ is larger than the magnitude of $\fscslab_{\bsame}$, we observe a change from attraction to repulsion even for relatively small patches. Upon further decreasing $\beta_2$, the radial component of the scaling function for the force reaches a maximum. This occurs if the overlap of the two patches is small. A further decrease of $\beta_2$ reduces $\fsc_r^{\left(2\right)}$. This can be understood by noting the fact that moving the patch on the second particle upwards reduces the area of the projection of the patch. Reducing $\beta_2$ even further moves the patch to the right hemisphere of the second colloid, where it does not participate in the mutual interaction between particles. This occurs in a region around $\beta_2=-90\degree$, where the radial component is constant.

Like in the previous case, we have observed several singularities of $\fsc_r^{\left(2\right)}$. Around the minimum at $\beta_2=90\degree$, if the patches face each other, there is a singularity of type \singi and the scaling function for the force exhibits a `V' shaped cusp with an opening angle which increases upon increasing $\thp$. In the case of Janus particles ($\thp=90\degree$) the opening angle reaches $180\degree$. This property of the force follows directly from the Derjaguin approximation: If $\beta_2$ is slightly shifted away from $90\degree$, around the circumference of the patches a region $\Lambdaopposite$ emerges. If the size of the patches increases, so does $\prdist$ within this region. Additionally, we have noticed singularities $\propto \left(\beta_2-\beta_{2,\text{sing.}}^{\left(i\right)}\right)^{3/2}$ of type \singii which occur if the patches are tangent. In Fig.~\ref{secK:compare1} these points are marked with circles.

Finally, we comment on the nature of singularities of the radial component of the scaling function for the critical Casimir force. As we have reported above, the nonanalyticities of $\fsc_r^{\left(i\right)}$ are similar to those of the scaling function $\psc$ for the potential rather than to nonanalyticities of the scaling functions for other components of the forces. On one hand this can be explained as a simple consequence of the similarity of the formulae from which $\fsc_r^{\left(i\right)}$  and $\psc$ are calculated (see Eqs.~\eqref{secC:DerjaguinForce} and~\eqref{secC:DerjaguinPotential}). On the other hand, all singularities of $\psc$ discussed in Sec.~\ref{secD:nonanalycities} manifest themselves upon changing the rotational configuration $\sphconf$ of the system. In contrast to all other components of the scaling function for the forces, the calculation of $\fsc_r^{\left(i\right)}$ does not require derivatives with respect to angles (see Eq.~\eqref{secV:forcetorque}); therefore the character of the nonanalyticities remains unchanged.

\subsection{Critical Casimir potential}

\begin{figure}
\includegraphics[width=0.45\textwidth]{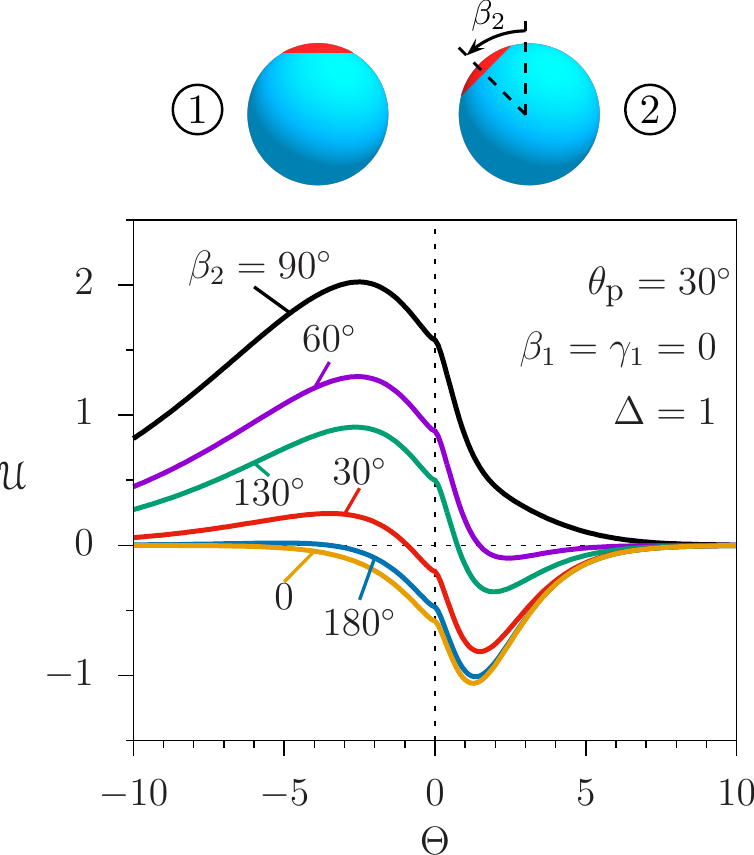} 
\caption{Scaling function $\psc$ of the critical Casimir potential as function of $\Theta$
for fixed $\Delta=1$, $\beta_1=\gamma_1=0$ and $\thp=30\degree$, for various values of $\beta_2$.
The configurations and the color code are the same as in Fig.~\ref{secK:Fsc_r_vs_Theta}.}
\label{secK:Psc_r_vs_Theta}
\end{figure}

\begin{figure*}
\includegraphics[width=0.99\textwidth]{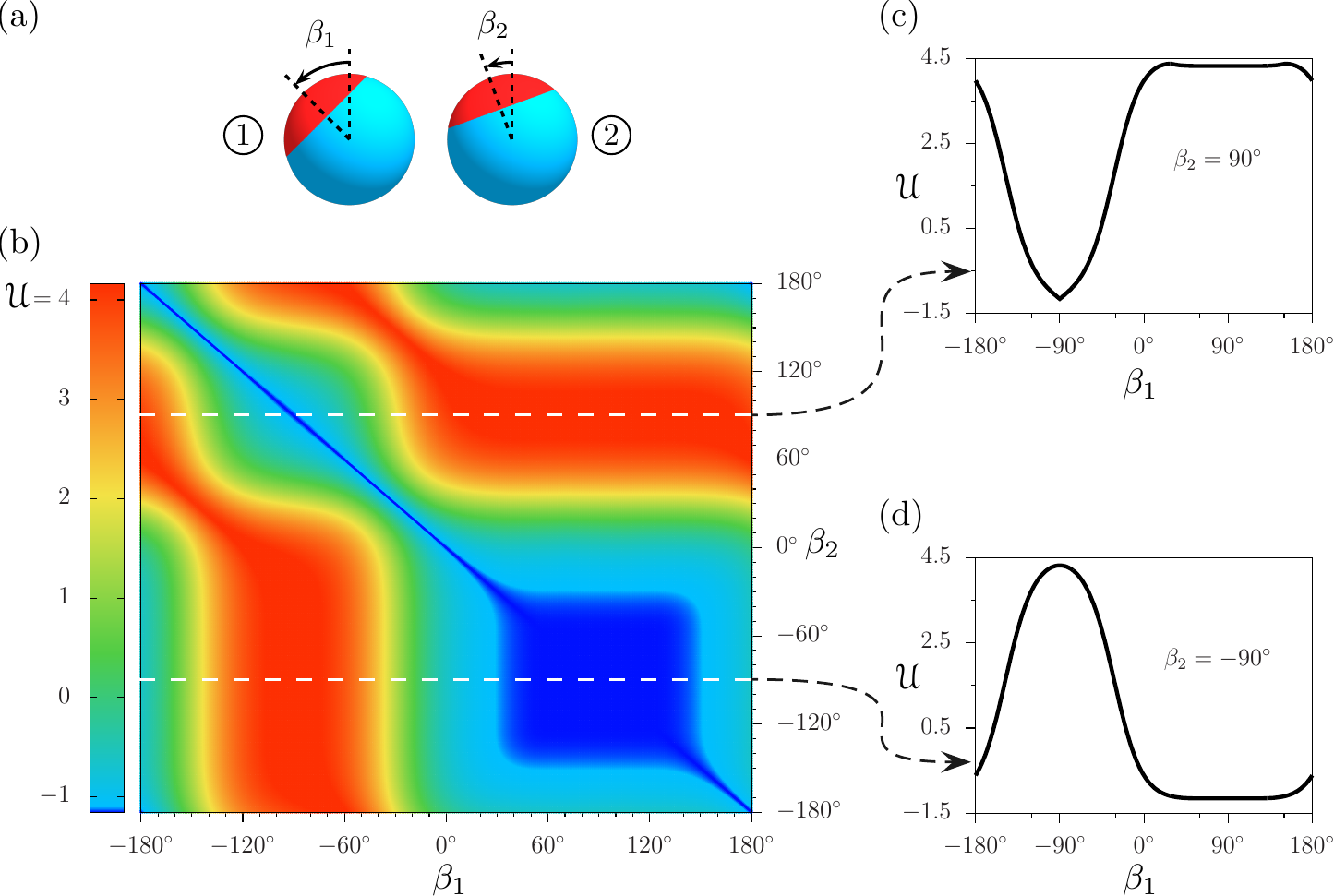} 
\caption{Scaling function $\psc$ of the critical Casimir potential for fixed values of $\Delta=\Theta=0.5$, $\gamma_1=0$, and $\thp=60\degree$ as a function of $\beta_1$ and $\beta_2$. (a) Schematic plot of the configuration of the colloidal particles. (b) Dependence of $\psc$ on $\beta_1$ and $\beta_2$; the value of the scaling function is indicated by the color code. In addition, we present two cuts of $\psc$ as functions of $\beta_1$ for fixed $\beta_2=90\degree$ (panel (c)) and $\beta_2=-90\degree$ (panel (d)), respectively. In panel (b) the configurations in the plots in panels (c) and (d) are marked by the white dashed horizontal lines.}
\label{secK:landscape-60-line}
\end{figure*}

In this subsection we present our results for the scaling function $\psc$ of the critical Casimir potential (see Eq.~\eqref{secC:potscaling}) calculated within the Derjaguin approximation.

In Fig.~\ref{secK:Psc_r_vs_Theta} the scaling function $\psc$ as a function of $\Theta$ is shown for various values of $\beta_2$ with all other parameters fixed ($\Delta=1$, $\beta_1=\gamma_1=0$, $\thp=30\degree$). If $\beta_2=0$ the patches on both spheres are in the topmost position and, within the Derjaguin approximation, the boundary conditions are the same everywhere. Thus for any value of $\Theta$ the function $\psc$ attains its smallest value via this configuration. Upon increasing $\beta_2$, with all the other parameters fixed, a region $\Lambdamp$ emerges (i.e., it becomes a nonempty set) and, as a result, $\psc$ increases. This growth is more pronounced for $\Theta<0$ (where the scaling function $\pscslab_{\bopposite}$ relevant for the slab geometry is the largest), and for $\beta_2>0$ the scaling function as a function of $\Theta$ has a maximum at $\Theta<0$. Correspondingly, upon increasing $\beta_2$ the minimum, visible in Fig.~\ref{secK:Psc_r_vs_Theta} for $\Theta>0$, becomes less deep, more shallow, and, eventually, somewhere between $\beta_2=60\degree$ and $\beta_2=90\degree$, it disappears. The scaling function reaches its maximum for $\beta_2=90\degree$. In this special configuration the area of $\Lambdaopposite$ is largest and, concomitantly, the mean value of the distance $\prdist$, between pairs of points on the surfaces projected onto the region, is smallest. A further increase of $\beta_2$ reduces $\psc$ and drives it negative again. For $\beta_2=180\degree$ the potential is only slightly larger than the lower bound corresponding to $\beta_2=0$. Finally, we note that the reported behavior of the scaling function $\psc$ strongly depends on the values of the fixed parameters. If the size $\thp$ of the patch is small and the scaled distance $\Delta$ is sufficiently large, $\psc$ can be negative for all values of $\Theta$ and $\beta_2$.

In Fig.~\ref{secK:landscape-60-line} we present the plot of the scaling function for the critical Casimir potential $\psc$ for fixed values of $\Delta=\Theta=0.5$, $\gamma_1=0$, and $\thp=60\degree$. The value of the function for the whole ranges of $\beta_1$ and $\beta_2$ is presented by resorting to the color code. We note that, due to reflection symmetry, the scaling function in Fig.~\ref{secK:landscape-60-line} is invariant under the transformations 
\begin{align}
\nonumber\left(\beta_1,\beta_2\right)&\mapsto \left(-\beta_2,-\beta_1\right)\quad \text{and}\\  \left(\beta_1,\beta_2\right)&\mapsto \left(180\degree-\beta_1, 180\degree-\beta_2\right).
\label{secK:symmetry}
\end{align}

The minimum of the function $\psc$, with $\psc_{\text{min}}\approx -1.1552$, is attained (within the Derjaguin approximation) if the boundary conditions are `$++$' or `$--$' for every point of $\Lambda$. This occurs if the patches are in the mirror--symmetric configuration (along the diagonal line $\beta_1=-\beta_2$) or if both patches are on those hemispheres which do not participate in the interaction (rectangle $\thp<\beta_1<180\degree-\thp$ and $-180\degree+\thp<\beta_2<-\thp$) (see Fig.~\ref{secK:landscape-60-line}(b)). In the latter case, the size of the rectangular region depends on the size $\thp$ of the patch in that it shrinks upon increasing $\thp$ and disappears completely for $\thp=90\degree$. As can be inferred from the cuts in Figs.~\ref{secK:landscape-60-line}(c) and \ref{secK:landscape-60-line}(d), crossing the region in which $\psc$ is minimal always reveals nonanalyticities: `V' shaped cusps (i.e., a singularity of type \singi) for the line $\beta_1=-\beta_2$ (outside the rectangle) and of type \singiii for the edges of the rectangle. We note that all properties reported here have been obtained within the Derjaguin approximation and we expect them to be absent beyond this approximation.

The maximal value of the scaling function $\psc$, with $\psc_\text{max}\approx 4.3949$, is attained at four isolated points: $\left(\beta_1\approx 19\degree, \beta_2\approx 99\degree\right)$ and at three points which can be generated from this first one by exploiting the symmetries described in Eq.~\eqref{secK:symmetry}. At these four special points various influences counter each other. Changing $\beta_1$ or $\beta_2$ increases the area of $\Lambdapp$; it moves parts of the patch to the far side hemispheres, where it does not participate in the interaction; or it moves the patch up or down which reduces the area of $\Lambdaopposite$. Unlike the minima, the precise positions of the maxima depend sensitively on the choices of $\Delta$, $\Theta$, and $\thp$.

The dependence of $\psc$ on $\beta_1$ for fixed $\beta_2=90\degree$ is presented in Fig.~\ref{secK:landscape-60-line}(c). This plot is symmetric around $\beta_1=-90\degree$, where it has a minimum and a nonanalyticity of type \singi. In this special configuration the two patches are facing each other. Upon increasing $\beta_1$ from $-90\degree$, there emerges a region $\Lambdaopposite$ and, as a result, $\psc$ increases. For $\beta_1>-60\degree$, increasing $\beta_1$ moves some part of the patch on the first particle to the left hemisphere, where (within the Derjaguin approximation) it is not participating in the interaction. At first, this effect is not very pronounced, but, upon increasing $\beta_1$ from $-60\degree$, it becomes more relevant. First, the growth rate of $\psc$ is reduced and finally, for $\beta_1\approx 28\degree$, the effect becomes dominant and the scaling function starts to decrease. If $60\degree<\beta_1<120\degree$, the patch on the first particle is located fully on the left hemisphere and the function $\psc$ becomes constant. Since $\psc$ is symmetric around $\beta_1=-90\degree$, a further increase of $\beta_1$ does not yield any new phenomena.

If $\beta_2=-90\degree$, the patch on the second colloid is fully located on the right hemisphere, and it does not influence the interaction. The cut of the scaling function $\psc$ in this special case is presented in Fig.~\ref{secK:landscape-60-line}(d). There is a maximum for $\beta_1=-90\degree$, which is attained if the patch on the first particle is in the closest possible position to the second particle. In this configuration, changing $\beta_1$ increases the distance between the points on the patch of the first particle and the points on the surface of the second sphere. Simultaneously, it reduces the area of $\Lambdapm$ and thus the repulsion; accordingly the scaling function for the potential decreases. This decrease continues for $60\degree<\beta_1<120\degree$ until the patch on the first particle is fully located on the left hemisphere; within this range of values for $\beta_1$ the function $\psc$ is the same as in the case of homogeneous spheres and thus has a flat minimum.

\begin{figure}
\includegraphics[width=0.42\textwidth]{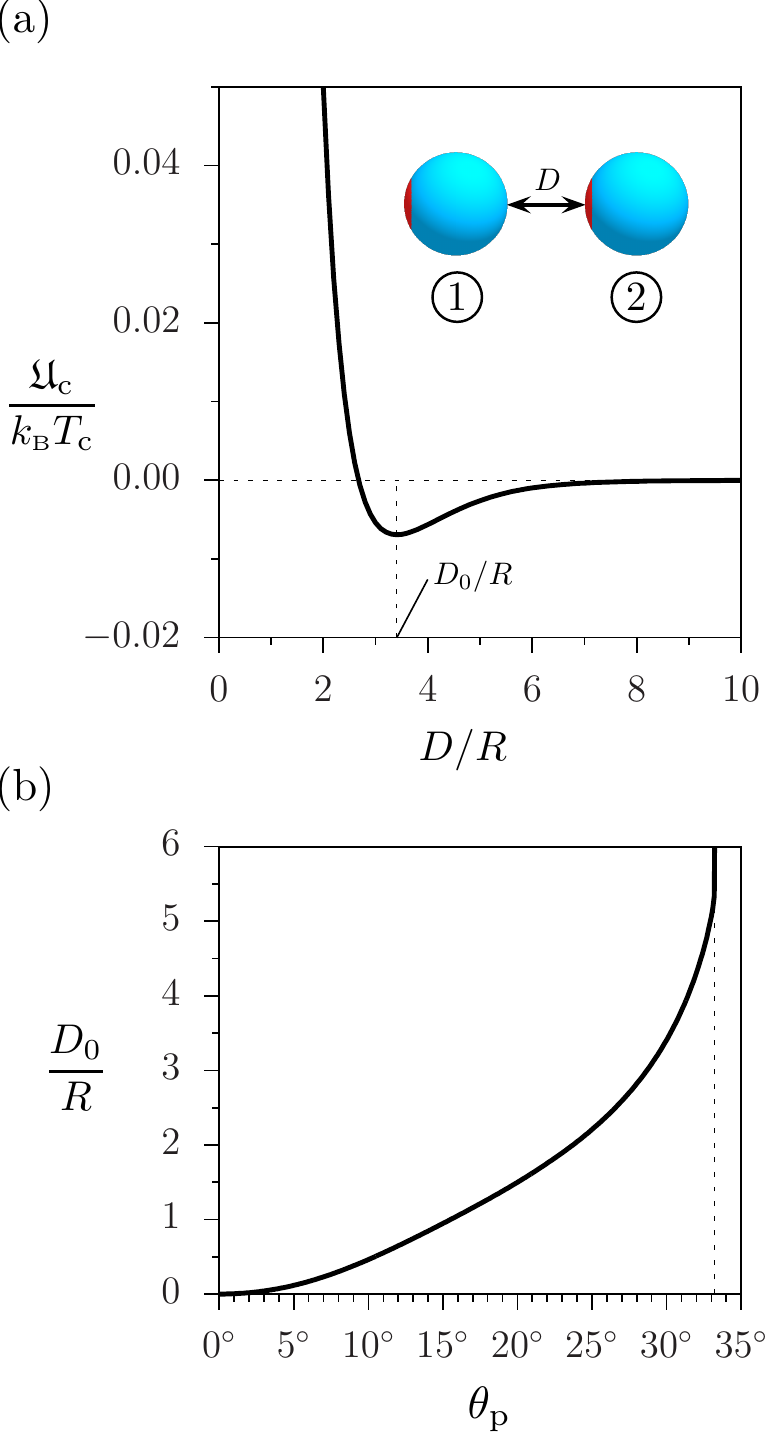} 
\caption{(a) Critical Casimir potential (in units of $\kB \Tc$, see Eq.~\eqref{secC:potscaling}) as function of the rescaled surface--to--surface distance $D$ between the particles. The size $\thp$ of the patch is $30\degree$ and the orientations of the colloids ($\beta_1=\beta_2=90\degree$ and $\gamma_1=0$) are presented schematically in the inset. The fixed temperature $T>\Tc$ is chosen such that $\xib=R$. (b) Position of the minimum of the potential as function of the size $\thp$ of the patch. $D_0\left(\thp\right)$ diverges for $\thp\to\thp^{\left(0\right)}$ with $\thp^{\left(0\right)}\approx 33.2\degree$ for the above choices of the parameters. The configuration of the particles and the temperature are the same as in panel (a).}\label{secK:min}
\end{figure}

Next, we study the dependence of the critical Casimir potential on the distance between the particles. In Fig.~\ref{secK:min}(a) we present the plot of the potential as a function of the surface--to--surface distance $D$ at a fixed supercritical temperature at which $\xib\left(T\right)=R$ for two particles with patches of size $\thp=30\degree$, both in the leftmost position (i.e., $\beta_1=\beta_2=90\degree$ and $\gamma_1=0$). Accordingly, the patch on the first particle is located on the left hemisphere and, thus, it does not influence the interaction. The potential has been calculated from the scaling function $\psc$ by using Eq.~\eqref{secC:potscaling}.

Within the Derjaguin approximation, the distance $\prdist$ between pairs of interacting points varies between $D$ and $D+2R$. Since the potential in the slab geometry increases rapidly upon decreasing the distance between walls (see Eq.~\eqref{secC:pscslabdef}), for $D\ll R$ the closest pairs of points (for which the boundary conditions are `$-+$') dominate the whole interaction. This explains why for small $D$ the potential is positive and diverges for $D\to 0$. If $D\gg R$, the effect of different distances for different pairs of points is negligible, and the sign of the potential is determined by the size of the patch. If the patch is sufficiently small, the area of $\Lambdamp$ is not large enough to facilitate a change of sign, so that for $D\to\infty$ the potential approaches $0$ from below. This leads to the conclusion that for small values of $\thp$ the potential as a function of $D$ has a minimum at a certain distance $D_0$. We expect that this holds also beyond the Derjaguin approximation.

In Fig.~\ref{secK:min}(b), we plot the dependence of the position of the minimum $D_0$ of the critical Casimir potential $\cpot$ (see Eq.~\eqref{secC:potscaling}) on the size $\thp$ of the patch. If the patch is very small, $D_0$ is close to $0$. Upon increasing $\thp$, the minimum becomes more shallow and moves towards larger values of $D$. Finally, when the size of the patch reaches a certain threshold value $\thp^{\left(0\right)}$, the position of the minimum diverges and the minimum disappears, so that beyond this threshold the potential is positive for all values of $D$. The precise value of $\thp^{\left(0\right)}$ depends on the ratios $\lim_{\omega\to\pm\infty}\pscslab_{\bsame}\left(\omega\right)/\pscslab_{\bopposite}\left(\omega\right)$, and thus it is very sensitive on how the Monte Carlo data for the slab geometry are extrapolated to large positive and negative values of the scaling variable $\omega$.

\begin{figure*}
\includegraphics[width=0.9\textwidth]{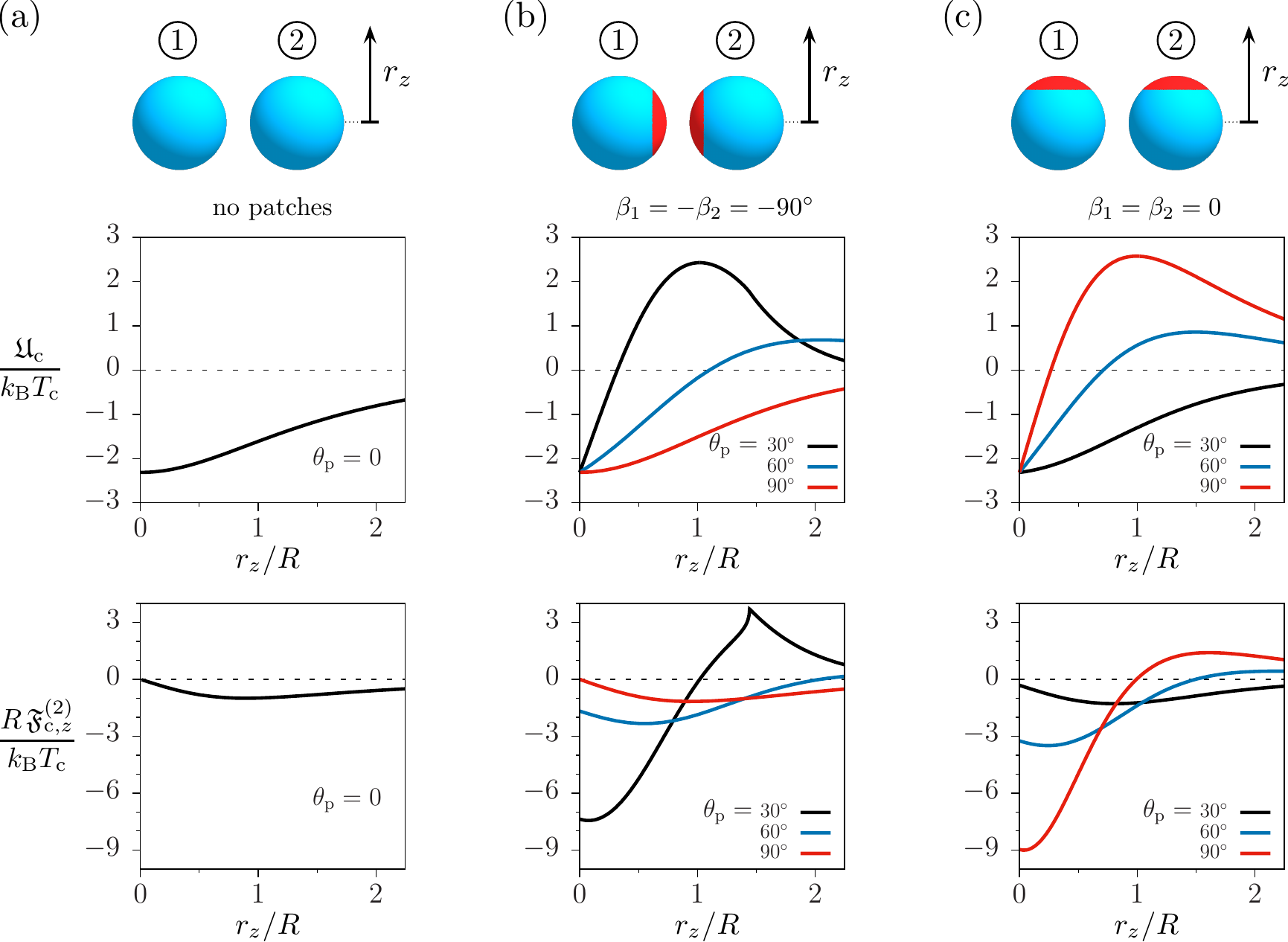} 
\caption{Critical Casimir interaction between two spherical particles of radius $R$, if one of them is moved in $z$ direction for (column (a)) homogeneous spheres, (column (b)) spheres with patches initially facing each other ($\beta_1=-90\degree$ and $\beta_2=90\degree$), and (column (c)) patches in the topmost position ($\beta_1=\beta_2=0$). For each configuration, the top graph presents the dependence of the potential on the shift $r_z$, while the bottom graph presents the $z$ component of the force acting on the second particle. In the columns (b) and (c) the curves for different sizes $\thp$ of the patches are plotted in different colors. For all plots the temperature $T>\Tc$ is chosen such that $\xib\left(T\right)=R$, and the initial distance between the colloids is $r_y=2.5\, R$ (i.e., initial surface--to--surface distance $D=0.5\,R$). In the columns (b) and (c) the interaction exhibits a singularity of type \singi at $r_z=0$ and of type \singii for $r_z=r_y \tan\thp$; the latter is beyond the range of the plots except for the case of $\thp=30\degree$ in column (b). We note that all the results have been obtained for systems in equilibrium; in this study we do not consider any dynamic effects.}
\label{secK:z-component}
\end{figure*}

We note that the minimum of the potential is observed only for shifting the second particle in radial direction. The configuration corresponding to the minimum is not stable with respect to rotations, and thus for $D=D_0$ the system occupies a saddle point.

\subsection{Non--radial components of the critical Casimir force}

In this subsection, we present our results for the critical Casimir force and potential, when the second particle is moved in $z$ direction (i.e.,~in a direction perpendicular to the line connecting the centers of the spheres in the initial position). For such a setup, even though the spheres are not rotated in the laboratory reference frame, their relative orientations change upon moving the particles. In the reference frame associated with $\sphconf^\ast$ (see Sec.~\ref{secB}) both spheres rotate around the $x$ axis and, at the same time, the distance between them increases. This way, the effect of rotations can be studied without actually performing any rotations, which makes the setup a good candidate for possible lattice--based simulations. We note that all calculations presented here have been carried out for a system in equilibrium; we do not consider any dynamic effects.

In Fig.~\ref{secK:z-component}, we plot the results for the potential and the $z$ component of the critical Casimir force acting on the second particle for various configurations and sizes of the patches. The results have been calculated from the scaling functions by using Eqs.~\eqref{secC:forcesf} and \eqref{secC:potscaling}. In all plots $r_x=0$, $r_y=2.5 R$, and the temperature is chosen such that $\xib\left(T\right)=R$ and $T>\Tc$.

We first consider spheres without any patches. In Fig.~\ref{secK:z-component}(a), the potential and the force for the case of homogeneous particles is plotted. The potential exhibits a minimum at $r_z=0$, which is the position where the colloids are in the closest possible position. Upon increasing $r_z$, the negative potential is gradually increasing towards $0$. If $r_z=0$, there is no $z$ component of the critical Casimir force. Upon increasing $r_z$ a force appears which acts in the negative $z$ direction. For small, increasing values of $r_z$ the absolute value of the $z$ component of the force increases and reaches its maximum value for $r_z\approx 0.9 R$, and, upon a further increase of $r_z$, it decays to $0$. This behavior can be understood by noting that the force acting between two homogeneous spheres is radial. For small values of $r_z$ the total force is large, but it is almost perpendicular to the $z$ direction, so that the $z$ component is small. For larger values of $r_z$, the angle between radial and $z$ direction decreases, but simultaneously the magnitude of the force decreases; the interplay between these two effects produces the maximum of the absolute value of the $z$ component of the force.

The critical Casimir potential and the $z$ component of the force in the presence of patches, initially located directly opposite to each other ($\beta_1=-90\degree$ and $\beta_2=90\degree$), are plotted in Fig.~\ref{secK:z-component}(b). If $r_z=0$, within the Derjaguin approximation, all pairs of interacting points have the same boundary conditions, and thus the value of the potential does not depend on the size of the patch. For $\thp>0$, in this configuration, there is a singularity of type \singi, i.e., the potential has a `V' shaped cusp around $r_z=0$; and the force is discontinuous, in the limit $r_z\to 0^+$ it is negative (acting downwards). In this limit, the magnitude of the force depends on the size of the patch. It is very small for $\thp\to 0$ and for $\thp\to 90\degree$, and maximal for $\thp \approx 20\degree$ (this behavior is not presented in the plot). Due to symmetry $\cforcez^{\left(2\right)}=0$ for $r_z=0$, which differs from the nonzero values in the limit $r_z\to 0^+$ (see the bottom panel in Fig.~\ref{secK:z-component}(b)). This observation is in full agreement with the properties of the Derjaguin approximation: the jump of the force at $r_z=0$ is generated by the region $\Lambdaopposite$ which emerges when the second colloid is moved slightly. For small shifts, this region has a ring--like shape around the circumference of the projections of the patches onto the projection plane. If $\thp$ is small, the circumference of the patch is very small, while for $\thp \lesssim 90\degree$ the region $\Lambdaopposite$ is located close to the boundary of $\Lambda$, where the surfaces of the spheres are almost perpendicular to the projection plane. In both cases the area of the region $\Lambdaopposite$ cannot be large, and thus, the jump of the force at $r_z=0$ is small.

Upon increasing $r_z$ from $0$, at first the magnitude of the $z$ component of the force increases. This is the same effect as the one occurring for homogeneous spheres: the radial component dominates the force, and the increase of $r_z$ reduces the projection angle. Upon a further increase of $r_z$, the influence of the patches becomes visible. The repulsion stemming from the region $\Lambdaopposite$ reduces the magnitude of the $z$ component of the critical Casimir force. If the size $\thp$ of the patch is sufficiently large, a further increase of $r_z$ changes the sign of $\cforcez^{\left(2\right)}$ so that the force starts to act upwards. If $r_z/r_y=\tan\thp$, i.e.,~if the projections of the edges of the patches on the two spheres are tangent, we observe a singularity of type \singii. (In Fig.~\ref{secK:z-component}(b) this occurs within the range of the plot only for $\thp=30\degree$.) For moderate values of $\thp$, the cusp with infinite slope in the plot of the force, which is characteristic for this type of singularity, coincides with the global maximum of the $z$ component of the force; for larger values of $\thp$, the maximum develops on the left side of the cusp. Upon a further increase of $r_z$, the force decays to zero as the distance between the patches grows.

In the third considered configuration both patches are in the uppermost position ($\beta_1=\beta_2=0$). The corresponding critical Casimir potential and the $z$ component of the force are plotted in Fig.~\ref{secK:z-component}(c). Like in the previous case, for $r_z=0$ the value of the potential does not depend on $\thp$ and there is a singularity of type \singi{}  --- the potential has a `V' shape around $r_z=0$, and the force jumps from positive values for $r_z<0$ to negative ones for $r_z>0$. In contrast to the previous case, here the absolute value of the force for $r_z\to0^+$ grows monotonically upon increasing the size of the patch. This observation can be understood from the fact that, if $r_z>0$ is very small, the typical distance $\prdist$ between the points in the region $\Lambdaopposite$ decreases upon increasing the patch size (for $\thp<90\degree$).

For small sizes of the patches the force is negative (i.e., it acts downward) for all values of $r_z$. If $\thp$ is sufficiently large, there appears a maximum where the force is positive (i.e., it acts upwards). Upon increasing $\thp$, at first, the maximum is very broad and it is located at very large values of $r_z$. A further increase of $\thp$ moves the maximum towards smaller values of $r_z$ and makes it more pronounced. In Fig.~\ref{secK:z-component}(c) the maximum is located within the range of the plot only for $\thp=90\degree$. This behavior can be understood on the basis of our formulae. For small values of $r_z$, the particles attract each other because the area of the region $\Lambdasame$ is much larger than the area of the region $\Lambdaopposite$. Upon increasing $r_z$, the area of the former region decreases, whereas the area of the latter region increases. If the patches are sufficiently large, this eventually produces the repulsion between the colloids, and the value of the $z$ component of the force becomes positive.

\subsection{Critical Casimir torque}\label{secK:CCT}

\begin{figure}
\includegraphics[width=0.45\textwidth]{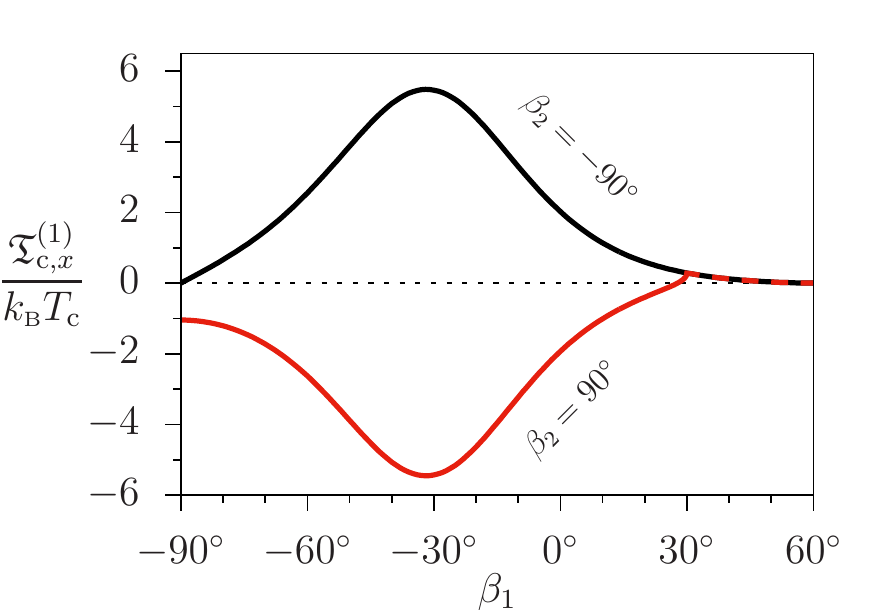} 
\caption{The $x$ component of the critical Casimir torque $\ctorquex^{\left(1\right)}$ (see Appendix \ref{secV}) acting on the first particle as a function of $\beta_1$ for $\beta_2=-90\degree$ (black curve) and $\beta_2=90\degree$ (red curve) with fixed $\Delta=0.5$, $\Theta=0.5$, $\gamma_{1}=0$, and $\thp=60\degree$. The scaling functions of the underlying critical Casimir potential for the two cases of $\thp=90\degree$ and $\thp=-90\degree$ are plotted in Figs.~\ref{secK:landscape-60-line}(c) and (d), respectively. In the configurations considered here there is no torque acting in $y$ or $z$ direction.}
\label{secK:torque_beta}
\end{figure}

In this subsection we discuss the critical Casimir torque acting on the colloidal particles. We calculate the scaling function for the torques $\tscv^{\left(1\right)}$ and $\tscv^{\left(2\right)}$ by using the numerical derivatives of $\psc$ and by deriving the torque via Eq.~\eqref{secC:torquesf}.

We first study the special case $\gamma_1=0$, in which the system is invariant under mirror reflection at the $yz$ plane. This symmetry implies that the torque can only act in  $x$ direction. Moreover, the system possesses the additional symmetry given by Eq.~\eqref{secK:symmetry} which implies
\begin{multline}
\tsc_x^{\left(2\right)}\left(\Delta, \Theta, \beta_1, \gamma_1=0, \beta_2\right)=\\
 -\tsc_x^{\left(1\right)}\left(\Delta, \Theta, -\beta_2, \gamma_1=0, -\beta_1\right), 
\end{multline}
so that it is sufficient to study the torque $\ctorquex^{\left(1\right)}$ acting on the first particle.

In Fig.~\ref{secK:torque_beta} the torque for $\thp=60\degree$, $\gamma=0$, and $\beta_2=-90\degree$, with fixed distance $D$, radius $R$, and temperature $T$, is plotted as function of $\beta_1$ as a black curve. In this configuration, the patch on the second particle is located on the right hemisphere and (within the Derjaguin approximation) it is not participating in the interaction; the torque is the same as if the second particle had no patch. We note that in this case the torque is proportional (with a negative coefficient) to the derivative of the scaling function for the potential $\psc$ with respect to $\beta_1$ (see~Eq.~\eqref{secV:torquex}); this function $\psc$ is plotted in Fig.~\ref{secK:landscape-60-line}(d).

For $-90\degree<\beta_1<60\degree$ the torque is positive, which means that in Fig.~\ref{secK:landscape-60-line}(a) it acts as to rotate the first particle anticlockwise. The maximum is located slightly below $\beta_1=-30\degree$. These properties are not surprising because increasing  $\beta_1$ (starting from $-90\degree$) adds points to the region $\Lambdapm$ with large values of $\prdist$ and removes those with smaller values of $\prdist$, which reduces the interaction free energy. The torque is largest close to the point, where the boundary conditions for the closest pair of points are changing; the neighborhood of these points is expected to dominate the interaction for small values of $D/R$. For $60\degree<\beta_1<120\degree$ the patch on the first particle is fully located on the left hemisphere, and thus in this interval there is no torque acting in the system.

The situation changes strongly for $\beta_2=90\degree$, i.e., when the patch on the second particle is moved to the leftmost position. In this case, the $x$ component of the critical Casimir torque is plotted in Fig.~\ref{secK:torque_beta} as a red curve, and the scaling function for the underlying potential is presented in Fig.~\ref{secK:landscape-60-line}(c). If $\beta_1=-90\degree$, the patches face each other and there is no torque. Upon a slight increase of $\beta_1$, the torque jumps from zero to a certain negative value (this jump is caused by a singularity of type \singi). Negative values of the torque mean that in Fig.~\ref{secK:landscape-60-line}(a) it acts as to rotate the first particle clockwise. Upon a further increase of $\beta_1$, the torque decreases, reaches its minimum for $\beta_1$ slightly below $30\degree$, increases, changes sign, and has a positive maximum for $\beta_1=30\degree$, where there is a singularity of type \singii with a characteristic cusp with infinite slope. For $\beta_1>30\degree$ (and $\beta_1<150\degree$), the torques for $\beta_2=90\degree$ and $\beta_2=-90\degree$ (red and black curve in Fig.~\ref{secK:torque_beta}, respectively) are identical.

The observed behavior follows directly from the properties of the Derjaguin approximation. For $\beta_1=-90\degree$, the boundary conditions are the same everywhere and the potential has a minimum. Upon increasing $\beta_1$ the region $\Lambdaopposite$ becomes a nonempty set, and thus the free energy increases. The minimum of the torque is located close to the point where the boundary conditions for the closest points change. Upon a further increase of $\beta_1$, a second effect appears according to which the patch on the first particle is moved to the left hemisphere and does not participate in the interaction. This increases the torque and eventually leads to the change of its sign. The maximum of the torque for $\beta_1=30\degree$ occurs at that position for which the region $\Lambdapp$ consists of a single point. For this configuration the area of $\Lambdaopposite$ is maximal. Finally, if $\beta_1>30\degree$, there is no overlap between patches, the region $\Lambdamp$ does not change with $\beta_1$, and thus the integral in Eq.~\eqref{secC:DerjaguinPotential} over this region does not depend on $\beta_1$ and does not contribute the derivative of $\psc$ with respect to $\beta_1$. Simultaneously, the integral over the region $\Lambdapm$ is exactly the same for $\beta_2=-90\degree$ and $\beta_2=90\degree$. This explains why the $x$ components of the torques are identical in these two cases (see $\beta_1>30\degree$ in Fig.~\ref{secK:torque_beta}).

\begin{figure}
\includegraphics[width=0.38\textwidth]{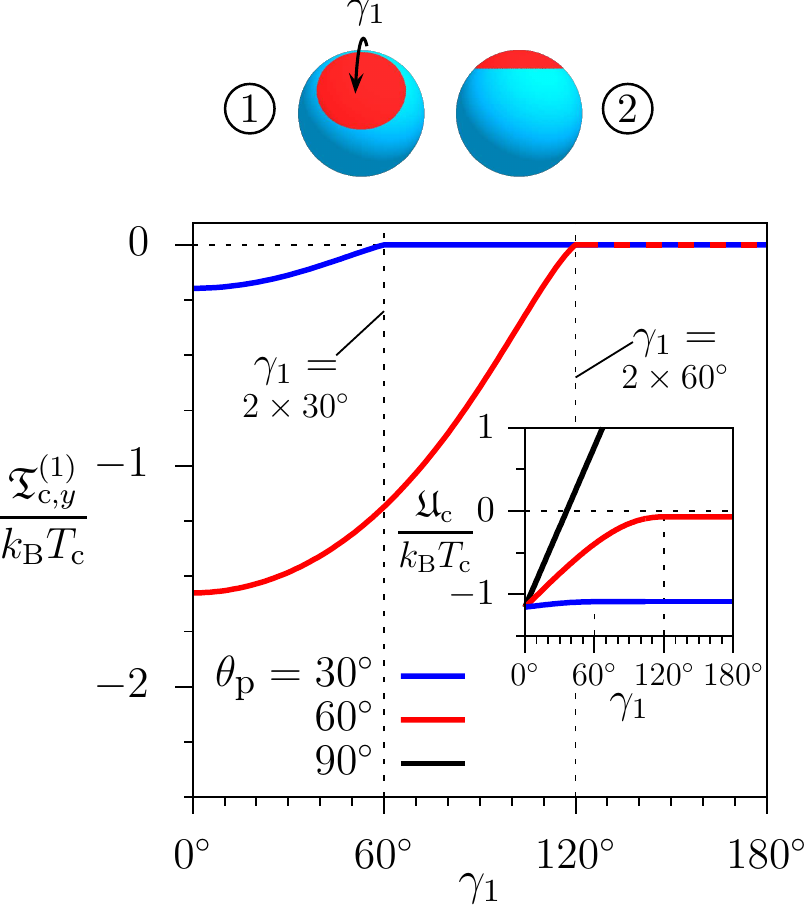} 
\caption{The $y$ component of the critical Casimir torque $\ctorquey^{\left(1\right)}$ (in units of $\kB \Tc$) acting on the first particle when it is rotated around the $y$ axis by an angle $\gamma_1$, as shown schematically above the plot. Here, $\Delta=\Theta=0.5$ and $\beta_1=\beta_2=0$. The curves for various sizes of the patches are marked by different colors. The inset presents the critical Casimir potential (in units of $\kB \Tc$) as a function of the rotation angle $\gamma_1$. For $\gamma_1>2\,\thp$ it is constant. In the case of $\thp=90\degree$ (black curve), for $0<\gamma_1<180\degree$ the potential $\cpot$ increases linearly so that the torque is constant ($\ctorquey^{\left(1\right)}\approx -3.68\, \kB \Tc$); this constant lies below the vertical range of the graph.}
\label{secK:pot_torque_gamma}
\end{figure}

We now discuss the case of $\gamma_1\neq 0$. For simplicity, we fix $\beta_1=\beta_2=0$ and $\Delta=\Theta=0.5$, and vary $\gamma_1$ from $0\degree$ to $180\degree$. In this case all components of the critical Casimir torque can now be nonzero. The dependence of $\ctorquey^{\left(1\right)}$ and $\cpot$ on $\gamma_1$ is presented in Fig.~\ref{secK:pot_torque_gamma}.

For $\gamma_1=0$, the patches on both particles face each other and, for any size $\thp$ of the patch, one has $\Lambda=\Lambdasame$; in this case and within the Derjaguin approximation, the critical Casimir potential attains its lowest possible value. Increasing the value of $\gamma_1$ renders the set $\Lambdaopposite$ nonempty, which increases $\cpot$ and generates a $y$ component of the torque acting against the rotation ($\ctorquey^{\left(1\right)}<0$). This situation changes if $\gamma_1$ exceeds the value of $2\,\thp$. In this case, there is no region $\Lambdapp$ anymore (i.e., this set is empty) and varying $\gamma_1$ shifts the region $\Lambdapm$, without changing its shape, in such a way that the integral in Eq.~\eqref{secC:DerjaguinPotential} remains the same. As a result, for $360\degree-2\,\thp>\gamma_1>2\,\thp$, the potential is constant and there is no torque in $y$ direction.

In the special case of Janus particles ($\thp=90\degree$), the edges of the patches form two diameters of the circle $\Lambda$ on the projection plane, and $\gamma_1$ is the angle between them. Since $\prdist$ depends only on the distance from the center of the circle $\Lambda$ on the projection plane, the integral in Eq.~\eqref{secC:DerjaguinPotential} is linear in the area of region $\Lambdaopposite$. As a result, the torque in $y$ direction is constant and negative for $0<\gamma_1<180\degree$, changes sign for $\gamma_1=0$ and $\gamma_1=180\degree$, and is constant and positive for $180\degree<\gamma_1<360\degree$.

We note that in the case of $\thp<90\degree$, for $\gamma_1=2\,\thp$ there is an unusual nonanalyticity. There is a jump in the second derivative of the potential with respect to $\gamma_1$ from a certain negative value for $\gamma_1<2\,\thp$ to $0$ for $\gamma_1>2\,\thp$. For $\gamma_1=2\,\thp$, there is just one point in $\Lambdapp$, but it is located at the boundary of $\Lambda$. This explains why the observed nonanalyticity cannot be classified into any of the three types described in Sec.~\ref{secD:nonanalycities}; we introduce a new type \singiv for such rare and untypical singularities (for further details see Appendix~\ref{secW:other}).

\subsection{Comparison with experimental results}
\label{secK:comparison-experiment}

\begin{figure}
\includegraphics[width=0.4\textwidth]{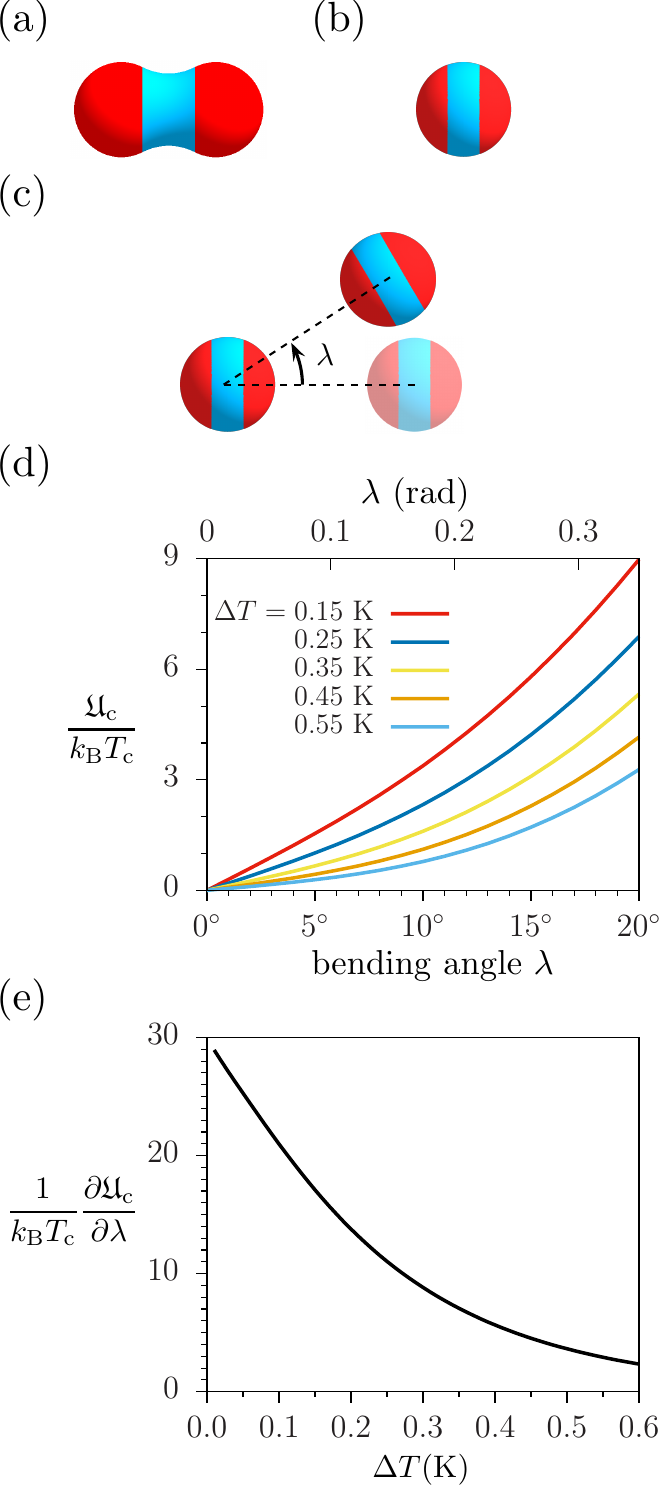}
\caption{(a) Schematic plot concerning the type of colloids studied experimentally in Ref.~\cite{Nguyen2017}. (b) Approximation of the colloid in panel (a) by a sphere, suitable to study within the present framework. (c) Schematic plot of the configuration of the particles used to calculate the bending stiffness of the interaction between two colloidal particles. A similar configuration has been considered experimentally (see Ref.~\cite{Nguyen2017} and the main text).
The particle with faded colors indicates the position of the right colloid prior to its angular displacement $\lambda$. (d) Present results for the critical Casimir potential as a function of the bending angle $\lambda=-\beta_1-90\degree$ for various values $T=\Tc-\Delta T$ of the temperature; the corresponding configuration of particles is presented in panel (c) and the size of the patches is chosen as $\thp=30\degree$. For $\lambda<90\degree-\thp$ the patches on the non--facing hemispheres of the colloids can, within the Derjaguin approximation, be neglected, and therefore the calculation of the potential can be carried out within the current, single patch framework. (e) First derivative of the potential with respect to $\lambda$ at $\lambda=0$ as function of $\Delta T$. This quantity determines the bending stiffness of the interaction between two colloidal particles.}
\label{secK:exp}
\end{figure}

One of the ways to test the validity of our calculation is to compare the results, obtained by using the Derjaguin approximation, with available experimental data. Up to our knowledge, most of the experimental papers focus on the problem of pattern formation by resorting to nonspherical colloids. Therefore, only a qualitative comparison is possible. In this subsection, we perform a simple check of our results with the pair potential measurements reported in Ref.~\cite{Nguyen2017}.

In this experiment a binary liquid mixture of heavy water and 3--methylpyridine close to its lower critical demixing point is used as a solvent. The colloidal particles, immersed in the mixture, have an oblong shape with the surface preferring 3--methylpyridine on the two tips and heavy water in the middle of the particle (see Fig.~\ref{secK:exp}(a)). For such a setup, the formation of superstructures by the colloids is investigated. Below the lower critical temperature (i.e., in the mixed region), depending on the concentration of the two components of the mixture, the colloids can align in two different ways: as needle--like chains, for which only the tips of two neighboring particles are in contact, and as fence--like chains in which the particles are parallel to each other (i.e., both the tips and the middle parts of neighboring particles are close to each other). Additionally, upon approaching the critical point a collapse of both types of chains is reported. Here, we are especially interested in the experimentally determined pair potential and bending stiffness of neighboring particles in needle--like chains.

Certainly, the actual particles as described above cannot be fully modeled within the present framework. Nonetheless, in order to proceed, we assume that the colloids can be approximated by a sphere of a radius $R$ with two chemical inhomogeneities, i.e., two circular patches of size $\thp$ in antipodal configuration (see, e.g., Fig.~\ref{secK:exp}(b)). The optimal radius $R$ and size $\thp$ of the patches are yet to be determined.

In Ref.~\cite{Nguyen2017} the effective pair potential of the interaction between two neighboring colloids in needle--like chains is estimated and inferred from the observed spatial distribution of the colloids. The radial component of the interaction is dominated by the competition between critical Casimir attraction and electrostatic repulsion. This effect has already been studied in the context of homogeneous spheres \cite{Stuij2017}. Therefore we can focus on the angular dependence of the effective pair potential, which is generated by the chemical inhomogeneity of the surface of the colloids and by their nonspherical shape. The dependence of the pair potential on the so--called bending angle (which is an analogue of the angle $\lambda$ in Fig.~\ref{secK:exp}(c)) turns out to be parabolic, and the bending stiffness (which is proportional to the second derivative of the potential with respect to the bending angle) decreases upon approaching the critical point (see~Fig.~2(c) in Ref.~\cite{Nguyen2017}).

In order to facilitate a comparison with our present results, it is necessary to determine the values of the relevant parameters in the experiment; most of them can be found in Ref.~\cite{Nguyen2017}: The critical temperature is $\Tc\approx38.55\degree\unit{C}$ and the bulk correlation length is $\xib\left(T\to\Tc\right)= \xi_0^\pm\left|t\right|^{-\nu},$ where $t=\left(\Tc-T\right)/\Tc$ is the reduced temperature, the critical exponent $\nu$ is given by Eq.~\eqref{secC:ampratio}, and $\xi_0^+\approx 1.6\unit{nm}$ has been estimated in the course of the experiment. (For a molecular liquid the value of $\xi_0^+$ is rather large and comparable to the one in nonionic micelle solutions \cite{Nguyen2017, Martinez2017}.) Since the mean value of the surface--to--surface distance $D$ between the colloids is not reported (most probably it was fluctuating during the measurement), we choose $D=0.2\unit{\mu m}$, for which the total radial potential (electrostatic plus critical Casimir) is found to exhibit a minimum (see~Fig.~2(a) in Ref.~\cite{Nguyen2017}). Because in the experiment the particles are not spherical, we estimate the effective radius to be $R=1.15\unit{\mu m}$. This value is certainly smaller than the actual size of the particles; however, for this effective radius the curvatures of the surface at the closest points of the two particles --- within the Derjaguin approximation this region provides the most important contribution to the interaction --- are the same as in the experiment. The only parameter which cannot be reliably estimated is the effective size $\thp$ of the patches; we have checked that in order to obtain the same order of magnitude of the potential as in Ref.~\cite{Nguyen2017}, $\thp$ should be roughly between $30\degree$ and $40\degree$.

We assume that the bending potential is generated solely by the critical Casimir interaction, neglecting a possibly inhomogeneous distribution of the surface charge. Accordingly, fixing the surface--to--surface distance to its experimental equilibrium value allows us in the following to not consider the electrostatic repulsion.

In order to calculate the effective potential, we put two colloids in the configuration shown in Fig.~\ref{secK:exp}(c). If $\lambda<90\degree-\thp$, one of the patches on each sphere is located on the distant hemisphere and thus, within the Derjaguin approximation, it does not participate in the mutual interaction. This implies that the potential can be calculated without introducing any modifications into our single patch framework (as described in Sec.~\ref{secB}). It is sufficient to consider spherical particles with a single patch of size $\thp$ in a configuration defined by $\beta_1^\ast=-90\degree-\lambda$, $\beta_2^\ast=90\degree$, and $\gamma_1^\ast=0$.
 
We note that the measurements were done in the vicinity of the lower critical demixing point, for which the mixed phase is observed below the critical temperature. Therefore, here the amplitude $\xi_0^+$ of the singular part of the correlation length is associated with $T<\Tc$.

In Fig.~\ref{secK:exp}(d) the dependence of the critical Casimir potential on the bending angle $\lambda$ is plotted for various temperatures below $\Tc$ (i.e., in the disordered phase) for a patch size $\thp=30\degree$. This is our analogue of the experimental data presented in Fig.~2(c) in Ref.~\cite{Nguyen2017}. In both systems the potential becomes stronger upon increasing the bending angle $\lambda$; shifting the temperature away from $\Tc$ reduces the growth of the bending potential as function of $\lambda$. This implies that $\lambda=0$ is the stable configuration with respect to rotations, and that the bending stiffness grows upon lowering the temperature below the critical point (i.e., into the disordered phase). In contrast to the experimental results, the calculated potential is not parabolic around $\lambda=0$. Instead, there is a singularity of type \singi and the function displays a `V' shape. As discussed in Sec.~\ref{secE}, the latter is an artifact of the Derjaguin approximation. Therefore, in our case the limit of the derivative $\lim_{\lambda\to 0^+}\partial \cpot/\partial \lambda$ is a suitable expression for the bending stiffness; we present a plot of this quantity in Fig.~\ref{secK:exp}(e). As in the experiment (see~the inset in Fig.~2(c) in Ref.~\cite{Nguyen2017}), the bending stiffness decreases upon shifting the temperature away from the critical point. We note that, because of the singularity at $\lambda=0$, our definition of the bending stiffness differs significantly from the bending stiffness used in Ref.~\cite{Nguyen2017} and therefore a detailed quantitative comparison is not possible.

The analysis presented above allows us to conclude that for the system under consideration the calculated properties are in qualitative agreement with the experimental data. The observed differences can be traced back to artifacts of the Derjaguin approximation.

\section{Conclusions}\label{secU}

We have studied the critical Casimir interaction between two identical patchy colloidal particles immersed in a binary liquid mixture close to demixing. The surface of the spherical colloids is chemically inhomogeneous in that it prefers one component of the binary solvent everywhere, except for a circular patch of angular size $\thp$, where there is an affinity to the second component of the 
binary mixture (see Fig.~\ref{secB:fig2}). When the mixture is close to its critical demixing point, the critical Casimir interaction between colloids emerges. It is anisotropic due to the inhomogeneous surface of the colloids and can be tuned by varying the thermodynamic parameters of the binary liquid mixture.

We have introduced the angles which, together with the surface--to--surface distance $D$ and the radius $R$ of the spheres, describe the relative configuration of the two particles (see Fig.~\ref{secB:fig1}). Using these parameters, we have formulated the scaling laws for the critical Casimir potential, force, and torque (see Eqs.~\eqref{secC:FTscaling} and \eqref{secC:potscaling}). We have derived the formulae which allow one to calculate the scaling functions for the force and the torque from that of the potential (see Eq.~\eqref{secV:forcetorque}). This part of our study is very general; it can be applied to particles of arbitrary shapes and chemical surface patterns.

We have used the following method to calculate the aforementioned universal scaling functions: First, the scaling function for the critical Casimir interaction potential between two colloids has been calculated by applying the Derjaguin approximation. Within this approximation, the scaling function for a nonelemental geometry is expressed in terms of an integral over the relevant scaling functions for the system in the slab geometry (see Eq.~\eqref{secC:DerjaguinPotential}). Second, by numerical differentiation we have derived the universal scaling functions for all components of the forces and torques acting on both particles (see Sec.~\ref{secD}). The validity and accuracy of this approach have been discussed in Sec.~\ref{secE}. Among the artifacts of the Derjaguin approximation we have identified various nonanalyticities of the scaling functions constructed upon this approximation (see Sec.~\ref{secD:nonanalycities}); their detailed analysis is a prerequisite for accurate numerical calculations.

In Sec.~\ref{secK}, we have applied the above method of calculation in order to study the critical Casimir potential, force, and torque acting between two colloids for various configurations and values of the relevant parameters. The analysis has revealed a complex interaction which can be tuned in a controlled way: The radial component of the force can be either repulsive or attractive and there can be a stable (i.e., with respect to radial shifts) position for an arbitrary value of $D$ (see Fig.~\ref{secK:min}). If in the initial configuration one of the colloidal particles is shifted perpendicular to the radial direction, the magnitude of the force counteracting the shift can be tuned by changing the angular size $\thp$ of the patches and the orientation of the particles. Moreover, for large distances the force can change sign (see Fig.~\ref{secK:z-component}). By varying the size of the patch a similar tuning has been observed for the torque (see Fig.~\ref{secK:torque_beta}).

Finally, we have performed a comparison of our results for the angular dependence of the critical Casimir interaction with available experimental data~\cite{Nguyen2017}. Even though the dependence of the free energy on the bending angle differs quantitatively between theory and experiment (due to the singularities introduced by the Derjaguin approximation), the temperature dependence of the bending stiffness shows qualitative agreement (see Sec.~\ref{secK:comparison-experiment}).

In order to gain further insight into critical Casimir interactions between colloids with inhomogeneous surfaces, it is necessary to study analogous systems by using a variety of different techniques, such as Monte Carlo simulations or mean field theory. The comparison of such independent results with those presented here would allow one to determine the actual interaction more accurately. Moreover, the approach presented here allows one to extend our techniques to more complicated shapes of patches or to nonspherical colloids. 

The results reported in this paper can also be used to perform molecular dynamics simulations in order to study clustering phenomena for patchy particles. Unfortunately, our numerical routines for calculating forces and torques are not fast enough to be directly usable for this purpose. Nevertheless, with additional efforts it is possible to overcome this problem: The scaling function for the potential $\psc$ can be decomposed into its singular and its nonsingular part. Using the results from Appendix~\ref{secW}, it is possible to propose exact formulae for the singular part, whereas the nonsingular part can be expanded into a series of appropriate orthogonal basis functions. The details of this procedure are beyond the scope of the present study.



\appendix

\vspace*{1cm}

\section{Forces and torques as derivatives of the interaction potential}\label{secV}

In order to present the method of deriving the forces and torques we focus on the $x$ component $\fsc_{x}^{\left(1\right)}$ of the scaling function for the critical Casimir force acting on the first particle. According to the definition of the potential, the force acting on the first sphere is
\begin{equation}\label{secV:forcex}
 \vctr{e}_x\cdot\cforcev^{\left(1\right)}=-\lim_{\epsilon\to 0}\left[ \cpot\left(T,R,D_\epsilon	,\sphconf^\ast_\epsilon\right)-\cpot\left(T,R,D,\sphconf\right)\right]/\epsilon,
\end{equation}
where $\vctr{e}_x$ denotes the unit vector in the direction of the $x$ axis, and $\sphconf_\epsilon^\ast=\left(\alpha_{1,\epsilon}^\ast, \beta_{1,\epsilon}^\ast, \gamma_{1,\epsilon}^\ast, \alpha_{2,\epsilon}^\ast, \beta_{2,\epsilon}^\ast\right)$ and $D_\epsilon$ are the relative configuration and distance between the colloids, respectively, after the first particle is shifted by the vector $\vctr{w}=\epsilon \vctr{e}_x$. We note that in the process of deriving $\sphconf_\epsilon^\ast$ and $D_\epsilon$, one can restrict the calculation to terms linear in $\epsilon$.

When the first particle is shifted by the vector $\vctr{w}$, its center is no longer in the origin. In order to reset the system to the special configuration, we first translate the whole system by the vector $-\vctr{w}$. This way, the center of the first colloid is back at the origin, and the center of the second colloid is at the point $\left(-\epsilon,r,0\right)$. Next, we apply the rotation $\mathbb{T}$ which puts the center of the second particle back onto the $y$ axis. In general, $\mathbb{T}$ consists of a rotation around the $z$ axis by the angle $\zeta_{z,\epsilon}=-\arctan\left(\epsilon/r\right)=-\epsilon/r+\mathrm{O}\left(\epsilon^2\right)$ and a rotation around the $y$ axis by the angle $\zeta_{y,\epsilon}=\epsilon \rho+\mathrm{O}\left(\epsilon^2\right)$, where the inverse length $\rho$ is yet to be determined. The resulting matrix of the rotation is
\begin{equation}\label{secV:Tmatrix}
 \mathbb{T}=\begin{pmatrix}
                     1 & \epsilon/r & \epsilon\rho \\
                     -\epsilon/r & 1 & 0 \\
                     \epsilon\rho & 0 & 1 
                    \end{pmatrix}+\mathrm{O}\left(\epsilon^2\right).
\end{equation}
This transformation increases the distance between the centers of the colloids from $r$ to $\left(\epsilon^2+r^2\right)^{1/2}=r+\mathrm{O}\left(\epsilon^2\right)$. Thus, up to the order linear in $\epsilon$, the distance remains unchanged:
\begin{equation}\label{secV:Dtransf}
 D_\epsilon=D+\mathrm{O}\left(\epsilon^2\right).
\end{equation}

We now investigate how the rotation $\mathbb{T}$ is changing the relative rotational configuration $\sphconf^\ast$ of the colloids. By using Eq.~\eqref{secB:stdrotation} the new rotational configuration of the colloids can be written as
\begin{subequations}\label{secV:angles}
\begin{align}
 \mathbb{R}\left(\alpha_{1,\epsilon}^\ast, \beta_{1,\epsilon}^\ast, \gamma_{1,\epsilon}^\ast\right)&=\mathbb{T} \cdot \mathbb{R}\left(\alpha_1, \beta_1, \gamma_1\right)\\
 \intertext{and}
 \mathbb{R}\left(\alpha_{2,\epsilon}^\ast, \beta_{2,\epsilon}^\ast, 0\right)&=\mathbb{T} \cdot \mathbb{R}\left(\alpha_2, \beta_2, 0\right),\label{secV:anglesb}
\end{align}
\end{subequations}
where we have put $\gamma_{2,\epsilon}^\ast=0$. Comparing the last column of the matrices on the left-- and right--hand side of Eq.~\eqref{secV:anglesb} and thereby using Eqs.~\eqref{secV:Tmatrix} and \eqref{secB:stdrotation} we obtain
\begin{subequations}
\begin{align}
 0&=\epsilon\left(\rho\cos\beta_2-1/r\ \sin \beta_2\right),\\
 -\sin\beta_{2,\epsilon}^\ast&=-\sin\beta_2,\\
  \intertext{and}
 \cos\beta_{2,\epsilon}^\ast&=\cos\beta_2.
\end{align}
\end{subequations}
For $\beta_2\neq \pm 90\degree$, the only solution of the above equations is $\rho=\left(\tan\beta_2\right)/r$ and
\begin{subequations}\label{secV:anglechange}
\begin{equation}
 \beta_{2,\epsilon}^\ast=\beta_2 +\mathrm{O}\left(\epsilon^2\right).
\end{equation}
By comparing various elements of the matrices in Eq.~\eqref{secV:angles}, a straightforward calculation yields
\begin{align}
 \alpha_{1,\epsilon}^\ast &=\alpha_1-\frac{\epsilon\cos\gamma_1}{r\cos\beta_1}+\mathrm{O}\left(\epsilon^2\right),\\
 \beta_{1,\epsilon}^\ast &=\beta_1+\frac{\epsilon\sin\gamma_1}{r}+\mathrm{O}\left(\epsilon^2\right),\\
 \gamma_{1,\epsilon}^\ast &=\gamma_1-\frac{\cos \gamma_1\ \tan \beta_1-\tan\beta_2}{r}\,\epsilon+\mathrm{O}\left(\epsilon^2\right),\\
  \intertext{and}
 \alpha_{2,\epsilon}^\ast &=\alpha_2-\frac{\epsilon}{r \cos\beta_2}+\mathrm{O}\left(\epsilon^2\right),
\end{align}
\end{subequations}
where we have additionally assumed that $\beta_1\neq\pm 90\degree$. Inserting Eqs.~\eqref{secV:Dtransf} and \eqref{secV:anglechange} into Eq.~\eqref{secV:forcex} leads to 
\begin{multline}
 \vctr{e}_x\cdot\cforcev^{\left(1\right)}=\frac{\cos\gamma_1}{r \cos\beta_1}\,\frac{\partial \cpot}{\partial \alpha_1}-\frac{\sin\gamma_1}{r}\,\frac{\partial \cpot}{\partial \beta_1}\\
 +\frac{\cos\gamma_1\ \tan \beta_1-\tan\beta_2}{r}\,\frac{\partial \cpot}{\partial \gamma_1}+\frac{1}{r\cos\beta_2}\,\frac{\partial \cpot}{\partial \alpha_2}.
\end{multline}
Finally, using Eqs.~\eqref{secC:FTscaling} and \eqref{secC:potscaling}, after some algebra one obtains the relation between the scaling function for the potential and the scaling function for one of the components of force:
\begin{subequations}\label{secV:forcetorque}
\begin{multline}
 \fsc_{x}^{\left(1\right)}=\frac{\Delta}{2+\Delta}\Bigg[\frac{\cos\gamma_1}{\cos\beta_1}\,\frac{\partial \psc}{\partial \alpha_1}-\sin\gamma_1\,\frac{\partial \psc}{\partial \beta_1}\\
 +\left(\cos\gamma_1\ \tan \beta_1-\tan\beta_2\right)\,\frac{\partial \psc}{\partial \gamma_1}+\frac{1}{\cos\beta_2}\,\frac{\partial \psc}{\partial \alpha_2} \Bigg].
\end{multline}

The calculation above can be repeated for all components of the forces and torques. Here, we skip the details and provide only the final results:
\begin{align}
 \label{secV:FIy}\fsc_{y}^{\left(1\right)}&=-\psc+\Delta\,\frac{\partial \psc}{\partial \Delta}+\Theta\,\frac{\partial \psc}{\partial \Theta},\\
\nonumber \fsc_{z}^{\left(1\right)}&=-\frac{\Delta}{2+\Delta}\Bigg(\frac{\sin\gamma_1}{\cos\beta_1}\,\frac{\partial \psc}{\partial \alpha_1}+\cos\gamma_1\,\frac{\partial \psc}{\partial \beta_1}\\
 &+\sin\gamma_1\ \tan\beta_1\,\frac{\partial \psc}{\partial \gamma_1}+\frac{\partial \psc}{\partial \beta_2}\Bigg),\\
 \fscv^{\left(2\right)}&=-\fscv^{\left(1\right)},\\
\nonumber \tsc_{x}^{\left(1\right)}&=-\Delta\Bigg(\frac{\sin\gamma_1}{\cos\beta_1}\,\frac{\partial \psc}{\partial \alpha_1}+\cos\gamma_1\,\frac{\partial \psc}{\partial \beta_1}\label{secV:torquex}\\
 &+\sin\gamma_1\ \tan\beta_1\,\frac{\partial \psc}{\partial \gamma_1}\Bigg),\\
 \tsc_{y}^{\left(1\right)}&=-\Delta\,\frac{\partial \psc}{\partial \gamma_1},\\
\nonumber \tsc_{z}^{\left(1\right)}&=-\Delta\Bigg(\frac{\cos\gamma_1}{\cos\beta_1}\,\frac{\partial \psc}{\partial \alpha_1}-\sin\gamma_1 \,\frac{\partial \psc}{\partial \beta_1}\\
&+\cos\gamma_1\ \tan\beta_1\,\frac{\partial \psc}{\partial \gamma_1}\Bigg),\\
\label{secV:TIIx}\tsc_{x}^{\left(2\right)}&=-\Delta\,\frac{\partial \psc}{\partial \beta_2},\\
\tsc_{y}^{\left(2\right)}&=\Delta\,\frac{\partial \psc}{\partial \gamma_1},\\
 \intertext{and}
\tsc_{z}^{\left(2\right)}&=\Delta\left(\tan\beta_2\,\frac{\partial \psc}{\partial \gamma_1}-\frac{1}{\cos\beta_2}\,\frac{\partial \psc}{\partial \alpha_2}\right).
\end{align}
\end{subequations}

A straightforward calculation shows that Eq.~\eqref{secV:forcetorque} does satisfy the relations in Eq.~\eqref{secC:FTrelations}. Additionally, the radial component of the force, given by Eq.~\eqref{secV:FIy}, has been checked numerically to agree with the results obtained from the Derjaguin approximation for the force via Eq.~\eqref{secC:DerjaguinForce}.

The results for the force and the torque (Eq.~\eqref{secV:forcetorque}) are not valid in the special cases $\beta_1=\pm 90 \degree$ or $\beta_2=\pm 90\degree$, for which certain coefficients in Eq.~\eqref{secV:forcetorque} diverge. A careful investigation shows that in these cases the rotations around the $z$ and $y$ axes are not independent. Moreover, an infinitesimal rotation can lead to a non--infinitesimal change of $\sphconf_{\epsilon}^\ast$. Since in most cases it is sufficient to consider $\beta_1$ and $\beta_2$ to be close but not equal to $\pm90\degree$, we refrain from reporting the formulae for the force and the torque in these special cases.

\section{Scaling functions for the slab geometry}\label{secX}

In this appendix we discuss the formulae for the scaling functions $\fscslab_{\bsame}$ and $\fscslab_{\bopposite}$ for the critical Casimir force in the slab geometry with same and opposite boundary conditions, respectively. In spatial dimension $\spdim=3$ these functions are obtained by interpolating and extrapolating the data calculated numerically by using Monte Carlo simulations~\cite{Vasilyev2009a,*Vasilyev2009b} in such a way, that all the known properties of the scaling functions are fulfilled.

In our analysis we are using the formulae obtained by A.~Gambassi et al.~\cite{Gambassi2019, Gambassi2009a}. They were successfully applied in several distinct studies \cite{Hertlein2008, Troendle2010, Labbe2016} but, the fitted functions have not yet been documented in the literature.

The fit uses a scaling law for the critical Casimir force which differs slightly from Eq.~\eqref{secC:cforceslab}:
\begin{equation}
\cforce^\mathrm{slab}\left(L,T\right)/A=\frac{\kB \Tc}{L^3}\,\mathcal{P}_{\mathrm{s}}\left(x\right),
\end{equation}
where $\mathcal{P}_\mathrm{s}$ is the scaling function and where the scaling variable $x=\left(L/\xi_0^+\right)^{1/\nu} t$ is linear in the reduced temperature $t$. Thus the relation between the scaling functions $\mathcal{P}_\mathrm{s}$ and $\fscslab_\mathrm{s}$ is given by
\begin{equation}
\label{secX:GambassiFormula1}
 \fscslab_{\mathrm{s}}\left(\omega\right)=\begin{cases}
               \mathcal{P}_{\mathrm{s}}(\omega^{1/\nu}) & \text{for } \omega\geqslant0,\\
              \mathcal{P}_{\mathrm{s}}\left(-\left|\omega\middle/\ampratio\right|^{1/\nu}\right) & \text{for } \omega<0,\\
            \end{cases}
\end{equation}
where $\ampratio$ is the ratio of the critical amplitudes of the bulk correlation length (see Eq.~\eqref{secC:ampratio}). The fitted shapes of the scaling functions are as follows:
\begin{subequations}\label{secX:GambassiFormula}
\begin{widetext}
\begin{align}
\label{secX:GambassiFormula2}
 \mathcal{P}_{\bsame}\left(x\right)&=\begin{cases}
             -1.5\times\left(-x\right)^{3 \nu}\exp\left[-1.89\times \left(-x\right)^{\nu}\right],                      & \text{for }  x<-12.424,\\
             -8.3354\times\left(5.5225 - x\right)^{3\nu}\exp\left[-1.89\times\left(5.5225 - x\right)^\nu\right],   & \text{for }  -12.424\leqslant x<3,\\
             -2.70476\times\left(1.10236 + x\right)^{3\nu}\exp\left[-\left(3.27051 + x\right)^\nu\right],          & \text{for }  3\leqslant x<16.5737,\\
             -1.51\times x^{3\nu}\exp\left(-x^\nu\right),                                   & \text{for }  16.5737\leqslant x,
            \end{cases}
\intertext{and}
\label{secX:GambassiFormula3}
 \mathcal{P}_{\bopposite}\left(x\right)&=\begin{cases}
             1.2264\times\left(7.398823 - x\right)^{3\nu}\exp\left[-0.6176\times\left(7.3988 - x\right)^\nu\right],    & \text{for }  x < -2.14689,\\
             5.410\times\left(9.51489 + x\right)^{3\nu}\exp\left[-\left(9.62878 + x\right)^\nu\right],           & \text{for }  -2.14689\leqslant x<14.8399 ,\\
             1.82\times x^{3\nu}\exp\left(-x^\nu\right),                               & \text{for }   14.8399 \leqslant x,
            \end{cases}
\end{align}
\end{widetext}
\end{subequations}
where the numerical coefficients have been chosen to reproduce as good as possible the data marked as `(i)' in Figs.~9 and 10 in Ref.~\cite{Vasilyev2009a}. In the above formulae the critical index $\nu$ has been assumed to be $0.63$. The accuracy of this fit is discussed in Sec.~\ref{secD:sfslab}. We note that the above forms of $\mathcal{P}_{\bsame}$ and $\mathcal{P}_{\bopposite}$ exhibit the correct behaviors \cite{Krech1994} for $x\to \infty$, $x\to-\infty$, and $\left|x\right|\ll 1$; the functions are continuous but have a slight jump of the derivative at the gluing points.

\section{Nonanalytic points of the interaction potential}\label{secW}

In this appendix we discuss the properties of the scaling function for the critical Casimir potential $\psc$ close to the points where it is not analytic due to applying the Derjaguin approximation in its present form.

In order to classify all possible nonanalyticities, it is necessary to focus on the boundaries of the patches and how they map onto the projection plane discussed in Sec.~\ref{secC:Derjaguin}. There are three relevant curves: the projection of the boundary of the patches on the first and second particle, and the circumference of the circle $\Lambda$ on the projection plane. The nonanalyticities of $\psc\left(\Delta,\Theta,\Omega^\ast \right)$ appear in configurations for which at least two of these curves are tangent. Here we elaborate on the three types of singularities (type \singi, \singii, and \singiii) which emerge if two of the curves are tangent; other types of nonanalyticities, categorized as type \singiv, are very rare and we refrain from analyzing them.

\subsection{Singularity of type \singi}\label{secW:kink}

The singularity of type \singi, as described in Sec.~\ref{secD:nonanalycities}, is a characteristic `V' shape of the potential, which appears when two patches are in a mirror--symmetric configuration.

For our detailed study, it is useful to introduce the notation $J_\mathrm{s}$ for the integrand in Eq.~\eqref{secC:DerjaguinPotential}:
\begin{equation}\label{secW:J}
 J_\mathrm{s}\left(\theta,\phi\right)=\frac{\Delta \sin^2\theta\ \sin\phi}{\left[\Delta+2\left(1-\sin\theta\ \sin\phi\right)\right]^2}\,\pscslab_{\mathrm{s}\left(\theta,\phi\right)}\left(\spomega\right),
\end{equation}
where $\Delta$ and $\Theta$ are fixed, `$\mathrm{s}$' denotes the boundary conditions (same or opposite), $\pscslab_\mathrm{s}$ is defined via Eq.~\eqref{secC:pscslabdef}, and the argument $\spomega$ of the scaling function is given by Eq.~\eqref{secC:parametrization_omega}. The formula in Eq.~\eqref{secC:DerjaguinPotential} for the scaling function $\psc$ can be transformed into
\begin{equation}\label{secW:kinkUdecomposition}
\psc\left(\Delta, \Theta, \sphconf^\ast\right)= \int_\Lambda\dd\theta\dd\phi\ J_{\bsame}\left(\theta,\phi\right)+\int_{\Lambda_{\bopposite}}\dd\theta\dd\phi\ \delta J\left(\theta,\phi\right),
\end{equation}
where we have introduced 
\begin{equation}\label{secW:deltaJ}
\delta J\left(\theta,\phi\right)=J_{\bopposite}\left(\theta,\phi\right)-J_{\bsame}\left(\theta,\phi\right). 
\end{equation}
The sets $\Lambda$ and $\Lambda_{\bopposite}$, as defined in Sec.~\ref{secC:Derjaguin}, are parametrized by the spherical coordinates $0\leqslant \theta\leqslant\pi$ and $0\leqslant\phi\leqslant\pi$ on the first particle. In Eq.~\eqref{secW:kinkUdecomposition}, the first term on the right--hand side is equal to the scaling function for spheres without patches. Thus it depends neither on the relative configuration $\Omega^\ast$ nor on the size $\thp$ of the patches. Therefore, the nonanalyticity of the type \singi is produced by the second term in Eq.~\eqref{secW:kinkUdecomposition}.

If the patches on two spheres are exactly in a mirror--symmetric configuration, $\Lambda_{\bopposite}$ is an empty set. If one of the spheres is slightly rotated, $\Lambda_{\bopposite}$ has a ring--like shape of variable thickness, following the circumferences of the projection of the patches onto the projection plane. For a more quantitative study, we assume that initially the patches on the spheres are in a mirror--symmetric configuration, opposite to each other, and that one of the spheres is rotated (around a certain axis) by a small angle $\tmpvarii$. We denote as $\Lambda_{\bopposite}^{\tmpvarii}$ the region on the projection plane corresponding to opposing boundary conditions in the final configuration.

In the process of rotation, the circumference of the projection of the patch on the rotated sphere passes all points of $\Lambda_{\bopposite}^{\tmpvarii}$. The circumference can be smoothly parametrized by $\theta=\tilde{\theta}\left(u,v\right)$ and $\phi=\tilde{\phi}\left(u,v\right)$, where $u_0\leqslant u\leqslant u_1$ parametrizes the points of the circumference of the patch and $0\leqslant v\leqslant \tmpvarii$ measures the rotation angle. We note that before the rotation the circumferences of both patches are given by $\tilde{\theta}\left(u,0\right)$ and $\tilde{\phi}\left(u,0\right)$, while after the rotation the circumference of the patch on the rotating sphere is given by $\tilde{\theta}\left(u,\tmpvarii\right)$ and $\tilde{\phi}\left(u,\tmpvarii\right)$.

The second integral in Eq.~\eqref{secW:kinkUdecomposition} can now be rewritten as
\begin{multline}\label{secW:kinkexpansion}
 \int_{\Lambda_{\bopposite}}\!\!\dd\theta\dd\phi\ \delta J\left(\theta,\phi\right)=\!\!\int_{u_0}^{u_1}\!\!\!\dd u\!\int_0^{\tmpvarii}\!\! \dd v\ \delta J\left(\tilde{\theta}\left(u,v\right),\tilde{\phi}\left(u,v\right)\right)\\
 \times \left|\frac{\partial \tilde{\theta}}{\partial u}\frac{\partial \tilde{\phi}}{\partial v}-\frac{\partial \tilde{\theta}}{\partial v}\frac{\partial \tilde{\phi}}{\partial u}\right|\sign \tmpvarii =\left|\tmpvarii\right| \int_{u_0}^{u_1}\!\! \dd u\left|\frac{\partial \tilde{\theta}}{\partial u}\frac{\partial \tilde{\phi}}{\partial v}-\frac{\partial \tilde{\theta}}{\partial v}\frac{\partial \tilde{\phi}}{\partial u}\right|\\
 \times \delta J\left(\tilde{\theta}\left(u,0\right),\tilde{\phi}\left(u,0\right)\right)+\mathrm{O}\left[\left(\tmpvarii\right)^2\right],
\end{multline}
where we have changed the variables of integration from $\left(\theta,\phi\right)$ to $\left(u, v\right)$, and where we have expanded the result for $\left|\tmpvarii\right|\ll 1$.

By using Eq.~\eqref{secW:kinkexpansion}, the expression for the scaling function in Eq.~\eqref{secW:kinkUdecomposition} can be written as
\begin{equation}
 \psc\left(\Delta, \Theta, \sphconf^\ast\right)=C_1+C_2\left| \tmpvarii\right|+\mathrm{O}\left[\left(\tmpvarii\right)^2\right],
\end{equation}
where $C_1$ and $C_2$ are parameters which do not depend on $\tmpvarii$. This explains the observed characteristic `V' shape of the potential, and proves that the derivative of $\psc$ with respect to $\tmpvarii$ does not exist for $\tmpvarii=0$ and its left and right limits have the same absolute value but different signs.

\subsection{Singularity of type \singii}

The singularity of type \singii appears if on the projection plane the circumferences of the patches are tangent in a single point. In order to describe these situations, we have introduced the overlap angle $\zeta_\text{overlap}^{\singii}$ (see Eq.~\eqref{secD:overlapb}). In order to study the scaling function for the potential in more detail, we decompose the integral in Eq.~\eqref{secC:DerjaguinPotential}:
\begin{multline}\label{secW:bumpUdecomposition}
\psc\left(\Delta, \Theta, \sphconf^\ast\right)= \int_\Lambda \dd \theta \dd \phi\ J_{\bsame}\left(\theta,\phi\right)+\int_{\Lambda_1}\!\!\dd \theta \dd \phi \ \delta J\left(\theta,\phi\right)\\
+\int_{\Lambda_2}\!\!\dd \theta\dd\phi\ \delta J\left(\theta,\phi\right)-2\int_{\Lambdapp}\!\!\dd \theta\dd\phi\ \delta J\left(\theta,\phi\right),
\end{multline}
where we have introduced the sets $\Lambda_1=\Lambdapp\cup \Lambdapm$ and $\Lambda_2=\Lambdamp\cup\Lambdapp$ which are the images on the projection plane of the patch on the first and the second particle, respectively. In the above equation, the integrand  $J_{\bsame}$ is given by Eq.~\eqref{secW:J} and $\delta J$ by Eq.~\eqref{secW:deltaJ}. The variables $0\leqslant\theta\leqslant\pi$ and $0\leqslant\phi\leqslant\pi$ are the spherical coordinates on the first particle and parametrize the set $\Lambda$ on the projection plane.

The first term on right--hand side of Eq.~\eqref{secW:bumpUdecomposition} describes the interaction of two spheres without patches, the second and the third term depend only on the position of the patch on one of the spheres. Since the nonanalyticity of type \singii depends on the position of both patches (via the overlap angle $\zeta_\text{overlap}^{\singii}$), it can only appear in the last term on the right--hand side of Eq.~\eqref{secW:bumpUdecomposition}:
\begin{equation}\label{secW:bumpterm}
 \psc_{\singii}=-2\int_{\Lambdapp}\!\!\dd\theta\dd\phi\ \delta J\left(\theta,\phi\right).
\end{equation}

If $\zeta_\text{overlap}^{\singii}<0$, the patches do not overlap, $\Lambdapp$ is the empty set, and $\psc_{\singii}=0$. For $0<\zeta_\text{overlap}^{\singii}\ll 1$, the area of overlap is small and the integral in Eq.~\eqref{secW:bumpterm} can be approximated by
\begin{equation}\label{secW:bumppart}
 \psc_{\singii}\approx-2 \delta J\left(\theta_0,\phi_0\right) \int_{\Lambdapp}\!\!\dd \theta \dd\phi,
\end{equation}
where $\theta_0$ and $\phi_0$ are the coordinates on the projection plane of the point at which the projections of both patches are tangent for $\zeta_\text{overlap}^{\singii}=0$. After some calculation, one obtains
\begin{multline}\label{secW:bumpaenergy}
 \psc_{\singii}=-\frac{8}{3\sin\theta_0}\sqrt{\tan \thp}\ \delta J\left(\theta_0, \phi_0\right) \left(\zeta_\text{overlap}^{\singii} \right)^{3/2}\\
 +\mathrm{O}\left[\left(\zeta_\text{overlap}^{\singii}\right)^{5/2}\right].
\end{multline}
By using Eqs.~\eqref{secW:bumpUdecomposition} and \eqref{secW:bumpaenergy}, for $\left|\zeta_\text{overlap}^{\singii}\right|\ll 1$ one finds
\begin{multline}
 \psc\left(\Delta, \Theta, \sphconf^\ast\right)=C_3 +C_4 \zeta_\text{overlap}^{\singii}\\
 +C_5 \left(\zeta_\text{overlap}^{\singii}\right)^{3/2}\Heaviside\left(\zeta_\text{overlap}^{\singii}\right)+\mathrm{O}\left[\left(\zeta_\text{overlap}^{\singii}\right)^2\right],
\end{multline}
which is in full agreement with Eq.~\eqref{secD:fbump}. In the above equation, $\Heaviside\left(x\right)$ is the Heaviside step function. $C_3$, $C_4$, and $C_5$ are parameters which do not depend on $\zeta_\text{overlap}^{\singii}$. The values of $C_3$ and $C_4$ are determined by first three terms in Eq.~\eqref{secW:bumpUdecomposition}, and
\begin{equation}
 C_5=-\frac{8}{3\sin \theta_0}\sqrt{\tan \thp}\ \delta J\left(\theta_0,\phi_0\right)
\end{equation}
is produced solely by the last term in Eq.~\eqref{secW:bumpUdecomposition}.

\subsection{Singularity of type \singiii}

The singularity of type \singiii occurs when the projection of the edge of one of the patches is tangent to the border of the circle $\Lambda$. This can happen in two different instances: (i) The patch is fully located on the hemisphere that is not participating in the interaction and, upon rotation, a part of it is moving towards the interacting hemisphere. (ii) The patch is fully located on the interacting hemisphere and the rotation moves a part of it to the other hemisphere.

Like in the previous case of the nonanalyticity of type \singii, the area of the relevant part of the patch is proportional to $\left(\zeta_{\text{overlap},i}^{\singiii}\right)^{3/2}$, where the overlap angle $\zeta_{\text{overlap},i}^{\singiii}$ is defined in Eq.~\eqref{secD:overlapSR}. It is possible to derive the expansion of the singular term in the interaction potential around the singularity. The result is very similar to Eq.~\eqref{secW:bumpaenergy}. Since for any of the limits $\theta, \phi \to 0,\pi$ we have $\delta J\left(\theta,\phi\right)/\sin\theta\to 0$, unlike the previous case, the leading order term in the expansion vanishes and, therefore, the next term, proportional to $\left(\zeta_{\text{overlap},i}^{\singiii}\right)^{5/2}$, dominates. Since this nonanalyticity is very weak, for our purposes a more detailed analysis is not necessary.

\subsection{Singularities of type \singiv}\label{secW:other}

All the other possible nonanalyticities require a special configuration according to which all three curves in the projection plane (i.e., the projection of the edges of the two patches and the circumference of the circle $\Lambda$) must meet in one point. In order to reach such a configuration, one has to tune all three angles, which makes these nonanalyticities rare. Moreover, the type of nonanalyticity detectable in the plots depends on how the parameters are changed. We note that, unlike for singularities of other types, the curves do not have to be tangent.

One of the possible situations, in which singularities of type \singiv become manifest, is presented in Fig.~\ref{secK:pot_torque_gamma}. In this case, the projections of the edges of the two patches are tangent at a point which is on the circumference of the circle $\Lambda$. As discussed in Sec.~\ref{secK:CCT}, in this case the scaling function for the potential as a function of $\gamma_1$ is continuous, has a continuous first derivative, but exhibits a jump in the second derivative.

A detailed analysis of all possible cases is quite complicated and, because all these nonanalyticities are artifacts of the Derjaguin approximation, goes beyond the scope of the present study.

%


\end{document}